\documentclass[12pt]{article}
\usepackage{placeins}
\usepackage{graphicx}
\def\begit{\begin{itemize}}
\def\endit{\end{itemize}}
\def\begc{\begin{center}}
\def\endc{\end{center}}
\def\begl{\begin{Large}}
\def\endl{\end{Large}}
\def\Begl{\begin{Large}}
\def\Endl{\end{Large}}

\title{History of ``Anomalous'' Atmospheric Neutrino Events: A First Person
Account\thanks{Based on a Talk presented at the Larry Sulak
Festschrift ``The Golden Age of Particle Physics and its Legacy'',
Boston University, October 21-22, 2005}
\thanks{The final publication is available at Springer via http://dx.doi.org/10.1007/s00016-016-0185-0}}
\author{\begl John M.~LoSecco \endl\\
\begl Physics Department \endl\\
\begl University of Notre Dame
\endl}
\date{\begl \today \endl}
\begin{document}
\maketitle
\begin{abstract}
The modern picture of the neutrino as a multiple mass
highly mixed neutral particle has emerged over 40
years of study.  Best known of the issues leading to this picture
was the apparent loss of neutrinos coming from the sun.
This article describes another piece of evidence that supports
the picture; the substantial reduction of high energy muon
type neutrinos observed in nature.  For much of the 40 year period,
before the modern picture emerged this observation was known as the
``atmospheric neutrino anomaly'', since as will be seen, these
neutrinos originate in the Earth's atmosphere.

This paper describes the discovery of the atmospheric neutrino anomaly.
We explore the scientific context and motivations in the late 1970's from which
this work emerged.  The
gradual awareness that the observations of atmospheric neutrinos were not as
expected took place in the 1983-1986 period.
\end{abstract}
\newpage
\tableofcontents
\listoffigures
\newpage
\section{Introduction}
Over the last 40 years the role of the neutrino in nature has been
studied and understood via observations of neutrinos from the sun,
extragalactic supernovae and cosmic rays.
The picture that has
emerged has been corroborated by observations of neutrinos from
nuclear reactors and particle accelerators.    The neutrino has
turned out to be a much more complicated physical system than
most elementary particles.  Many kinds of neutrinos and
antineutrinos are now known to exist and transformations among them
can now explain many of the odd features noted in the observations.

It is frequently convenient to label neutrinos by their properties under the
charged current weak interaction.  Under the influence of the charged current
the neutrino will turn into a charged particle, a lepton.  The kind of charged
particle produced by the interaction is then used to label the kind of
neutrino.  If a negatively charged electron, muon or tau is produced the
neutrino is regarded to be an electron, muon or tau neutrino.  An antineutrino
would produce a positively charged lepton.

Nuclear reactions in the sun produce electron neutrinos.  Cosmic ray
interactions in the Earth's atmosphere produce a mixture of muon and
electron neutrinos and antineutrinos.

This paper discusses the history of the discovery of what was known as the
``atmospheric neutrino anomaly''.  This effect is widely regarded as one of the
strongest bits of experimental evidence for neutrino oscillations and hence
a neutrino mass.

I will frequently use the terms neutrino mixing and
neutrino oscillations interchangeably since oscillations require mixing.
Neutrino mass {\em differences} are also required for the mixing to manifest
itself as a time and distance dependent variation in the neutrino properties,
an oscillation.

Atmospheric neutrinos are neutrinos, which are produced by cosmic ray
interactions in the Earth's atmosphere.  Cosmic rays interact strongly to
produce pions, kaons and other unstable particles by collision.  The decay
of these produced particles
yields muons, the most common component of cosmic rays at ground level.
The decays also produce neutrinos which, until the period involved
here, were very difficult to observe\cite{SouthAfr,Kolar}.

The atmospheric neutrino anomaly refers to the fact that the muon neutrino
flux of atmospheric neutrinos is substantially lower than the expected value.
It has many similarities to the solar neutrino problem, in that the neutrinos
were observed in nature and the observations were well below expectation.
The modern view is that the atmospheric neutrino anomaly and the solar
neutrino problem are closely related via the phenomena of three flavor neutrino
oscillations.

The time period of this paper is primarily 1978 to 1988.  Some issues of the
state of physics
prior to 1978 are reviewed to set the stage for the events that followed.  The
early and mid 1970's was a very productive period for particle physics.
Electro-weak unification, quantum chromodynamics, grand unified theories,
supersymmetry and their experimental underpinnings were all developed in this
period.  String theories also became well established during the 1978-1988 period
of our story.

\section{Scientific Context}
Among the significant issues in neutrino physics in the mid 1970's
were a number of ``discoveries'' which have since been resolved and were
ultimately not confirmed.  Scientific
discovery is rarely linear and this section describes reports that
could have been central to the issue of neutrino mixing, but many of
them were ultimately tangential.  The goal of this section is to put
the subsequent story into its historical context.

The weak neutral current is the interaction that permits neutrinos
to interact without changing into a charged lepton.  It is rarely
relevant in nuclear and particle decays since the much stronger
electromagnetic interaction can do these more rapidly.  The weak neutral
current was first observed in accelerator neutrino beams.

The discovery of the weak neutral current (which turned out to be true)
involved considerable uncertainty\cite{ANeutCur}.  This uncertainty has become
known in the field as ``alternating neutral currents'', since the
reports of discovery came and went and then returned.  In fairness to those
authors one should emphasize that doing a careful job is frequently
inconsistent with making a dramatic discovery.  To rush such checks can
lead to uncovering
evidence that both supports and refutes a conclusion.  As each piece of
evidence is analyzed the ``conclusions'' may change.  If one does many checks
the conclusions can change many times.

The high y anomaly\cite{High-y} was also a significant contributor to what
would
follow.  This disagreement of observed kinematic distributions (the $y$
variable, the fraction of the initial neutrino energy carried by the final state
muon) for (anti)neutrino
interactions at high energy with expectations, was taken by some as evidence
for right handed currents.  A neutrino mass would provide a natural source for
right handed currents.  Such currents helped motivate the possibility
of neutrino oscillations in the mid 1970's.  Subsequent experiments
failed to support the presence of the high y discrepancy.

Some evidence for the violation of the muon number and electron number
conservation was found in the decay $\mu \rightarrow e \gamma$.\cite{AnnRev86}
(The symbol $\mu$ represents the muon, $e$ the electron and $\gamma$ a
gamma ray.)  In such
a decay muon number decreases and electron number increases but the total
remains constant.
The concepts of muon number, electron number and tau number had been introduced
to explain the absence of transitions not forbidden by any other known
conservation law.  The concept declared that each of the leptons contained
a unique property that was conserved in all interactions.  The charged lepton
and its corresponding neutrino shared this property.  The concept of lepton
number explains, for example why the two neutrinos produced in muon decay
$\mu \rightarrow e \nu \bar{\nu}$ had to have different flavors.
The decay $\mu \rightarrow e \gamma$ is forbidden since both the muon number
and the electron number change by one unit.

The existence of the decay $\mu \rightarrow e \gamma$ would remove a
constraint that prevented neutrinos from mixing.
Ultimately the evidence for the decay $\mu \rightarrow e \gamma$ was not
confirmed.  Searches for this lepton number violating decay continue today.

Direct kinematic evidence for a neutrino mass was published by a Soviet group
under Lubimov\cite{Lubimov}.  Lubimov had used the classic method of studying
the high energy end of the tritium beta decay spectrum with a precision
spectrometer.  A neutrino mass would produce distortion of the end point
since a mass would limit the phase space for the highest energy electrons
from this beta decay.  (Their maximum kinetic energy would be lower since
some of the decay energy would appear as the neutrino mass.) At the time
the previous best upper bound on the
neutrino mass was 60 eV using a similar method\cite{Bergkvist}.
From 1980 onward there was unrefuted evidence from this
group for a neutrino mass of from 30-40 eV.  It took at least a decade of
effort to
eventually show this result was in error.  Currently there is no established
value for the neutrino mass, only upper limits.  Though observations do
support the existence of neutrino mass differences.

A novel method of searching for neutrino flavor transformations yielded some
evidence for oscillations in 1979\cite{Pasierb}.  The concept was a good one.
Use the neutral current interaction of neutrinos on deuterium to measure the
neutrino flux and use the
charged current interaction on the same deuterium to measure the electron
neutrino content at the same time.  The nuclear reactor source of neutrinos only
makes electron type antineutrinos.  These are of too low an energy to have
charged current interactions if they transform into muon or tau neutrinos.  But
the transformed neutrinos would still have neutral current interactions.  So the
neutral current observations measured the total neutrino content and the charged
current interactions measured only the electron type neutrinos.  If the
two measurements did not agree some of the electron neutrinos had been
transformed.
This in fact, was the method used by the Sudbury Neutrino
Observatory (SNO) to resolve the solar neutrino puzzle.  (The SNO group
used the charged current neutrino reaction on deuterium
$\nu_{e} D \rightarrow e^{-} p p$ to measure the electron neutrino content of
the solar flux.  They used the neutral current reaction
$\nu D \rightarrow \nu p n$ to measure the flux of all types of neutrinos
independently of the neutrino type.)  Careful checks
ultimately indicated that the conclusions of Pasierb {\em et al.}\cite{Pasierb}
that neutrinos had transformed were in error.

The period of the 1970's had many exciting neutrino observations which inspired
subsequent work.  Many of these including the high y anomaly,
$\mu \rightarrow e \gamma$, the Lubimov neutrino mass and the Pasierb
neutrino oscillations turned out not to hold up.  Others, such as the
weak neutral current, survived an ambiguous origin to become cornerstones
of modern particle physics.
\subsection{Solar Neutrinos}
While it is often assumed that the solar neutrino problem provided strong
motivation to study neutrino oscillations I think this is not true.  There
were three possible explanations for the solar neutrino problem, and until the
discovery of the Mikheyev-Smirnov\cite{MikSmir} effect in 1985, neutrino
oscillations was the
least popular.  Both the nuclear chemistry method used for the initial
observations and the obscure branch of the solar reaction model needed to
produce the energetic neutrinos needed for that detection method were also
suspect.  The Mikheyev-Smirnov\cite{MikSmir} effect, also known as MSW,
permitted
large neutrino flavor changes from small neutrino mixing due to contributions
of the neutrino electron charged current to the forward scattering amplitude.
When traversing matter electron type neutrinos had an additional interaction
with the electrons in the matter.  This interaction, under certain conditions,
can enhance neutrino oscillations. 
The MSW effect became better known in the west due to the work of
Bethe\cite{Bethe} in 1986.

\section{Inspiration}
A number of incidents were instrumental in focusing attention on the neutrino
oscillations\cite{Pontecorvo} question.  The discovery of the $\tau$
lepton\cite{Perl} clearly
indicated our knowledge of the lepton sector was incomplete.  Thinking at the
time was that such a charged lepton should be accompanied by a neutral
{\em massless} partner, the $\nu_{\tau}$.    The $\tau$ lepton was the
first clear evidence for a third family.

The presence of three distinct
neutrino types significantly expanded the phenomenology of neutrino
oscillations\cite{New-Osc}.  In particular the number of parameters to describe
neutrino oscillations would rise to 4 angles, from the one needed when only
two families were present.  The work of Kobayashi and Maskawa\cite{KobMask}
had also made it clear that there was now enough structure in the lepton
sector to permit the presence of CP violation, a very rare phenomenon at
the time.  (CP violation is a manifest difference in the properties
of matter and antimatter.
Kobayashi and Maskawa had pointed out that a physical theory
needed three particle families, if it was to have enough
degrees of freedom for the phenomena of CP violation to manifest itself.
They built upon the work of Cabibbo\cite{Cabibbo} who pointed out that
nuclear beta decay via the weak charged current could be understood as a
transition involving a superposition of the d quark, normally found in
protons and neutrons, with the much rarer strange quark.)
\begin{figure}
\parbox[b]{6.5in}{\includegraphics[height=6.5in,angle=270]{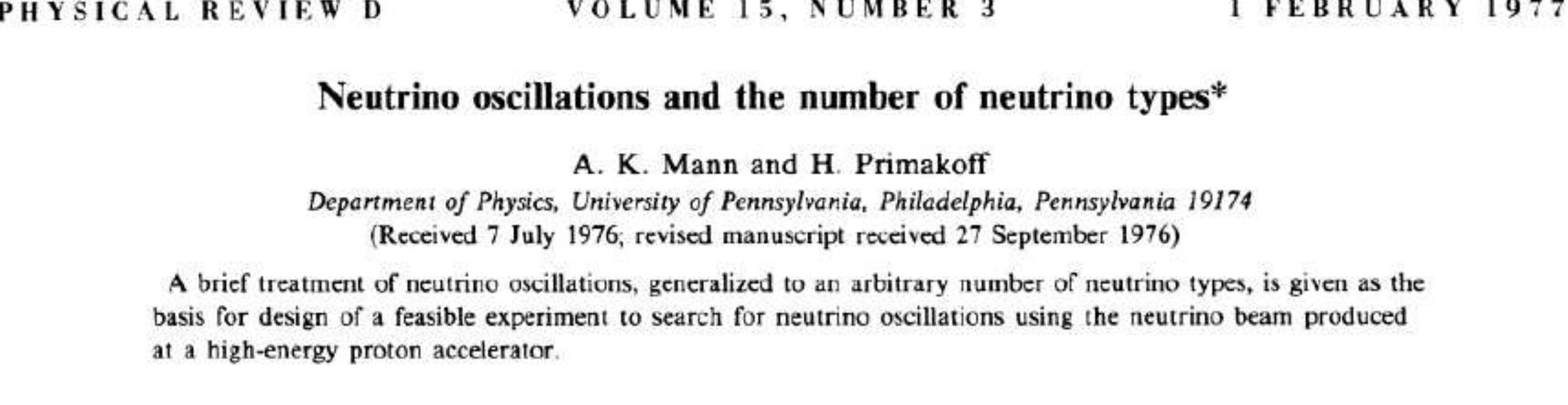}}
\caption{\label{fig:MP}Title and abstract of the 1976 Mann and Primakoff paper that
explored the possibilities of three neutrino oscillations}
\end{figure}
The Mann and Primakoff paper\cite{New-Osc} stimulated a good deal of research.
For example, considerations of the effect of bulk matter on neutrino
transport\cite{Wolf} were contemplated in response to the suggestion
in\cite{New-Osc} that one would want 1,000 kilometer long neutrino beams
slicing through a cord of the Earth.  A number of existing neutrino experiments
were modified to make them more accessible to observing neutrino oscillations.
Figure \ref{fig:MP} shows the title and abstract from this paper\cite{New-Osc}.

Another issue in this period that contributed to the interest in neutrino
oscillations was the general maturation of accelerator based neutrino physics.
What had been an exotic and difficult program to create tertiary neutrino beams
had come to fruition at most of the major high energy physics labs.  The
presence of such facilities made it much easier to take the next step in the
study of neutrino properties.  The interferometry method embodied in neutrino
oscillations gave one high energy accelerator based access to neutrino mass
scales of the order of electron volts.  Neutrino oscillations are essentially
an interferometric effect.  (Multiple amplitudes lead to the same state and the
amplitudes can interfere).  The initial neutrino flavor state is a particular
superposition of the neutrino mass states.  Since the different
neutrino mass states evolve in time at different rates an initially pure
flavor state will turn into a superposition of flavor states at a future time.
Measuring the flavor content as a function of time gives one access to both
the degree of neutrino mixing, via the magnitude of the impurity, and the
neutrino mass differences scale, by when the flavor variation
emerges and reaches a maximum in time.

\section{Inspiration 2}
A second source of inspiration came from the very rapid progress in
theoretical physics in the early 1970's.  Nonabellian gauge theories as part
of the standard model started to appear to play a role in nature.
The discovery of weak neutral currents provided experimental
support for electro-weak unification, in the context of a gauge theory.
Asymptotic freedom\cite{AsFr}, the weakening of the strong interactions
at high energy, the
dynamics underlying quantum chromodynamics (QCD) was discovered.  With the
running of the strong interaction coupling constant came the possibility that
at some momentum it would meet the electro-weak value and Grand Unified
theories\cite{GUTS} (GUTS) were created.  These theories combined three forces of
nature into one large group and demonstrated how the underlying distinctions
would emerge at normal energies.  At some high energy, called the unification
scale, all the interactions would be identical, with the same interaction
strength.  At lower energies the coupling constants and selection rules would
diverge to form, what appear to be three distinct forces of nature.

Such unification was not without cost.  Rather quickly it was realized that
some of the additional interactions present in Grand Unified theories would
lead to new, rare phenomena\cite{PDK}, such as the decay of the proton to
leptons and mesons.  The final state of proton decay would conserve electric
charge, angular
momentum or spin and energy but it would not conserve the values of baryon
number or lepton number.  The theory was not in conflict with observation
since the lifetime predicted
was still several orders of magnitude longer than experimental limits on proton
decay at the time.

The possibility of experimental confirmation of Grand Unified theories via the
observation of proton decay became a goal of late 1970's to early 1990's
particle physics and is still important today.  No convincing evidence has
been reported for proton instability.  Even now, the best we have are upper
limits on its lifetime.  The story of how technical problems were solved
to reduce costs such that massive detectors capable of making interesting
measurements of the proton lifetime is a long one that we can not discuss here.
A recent talk\cite{Cortez} outlines how the major design decisions were made
in the period 1978-1979 and the first detectors constructed by 1982.

Much of the work described in the current paper took place in the context
of a collaboration centered at the University of California at Irvine,
the University of Michigan and Brookhaven National Laboratory, known as IMB. 
The collaboration was formed in early 1979 to construct a massive deep
underground detector (eventually built near Cleveland, Ohio) to discover proton
decay.  The Irvine group included
many of the co-discoverers of atmospheric neutrinos.

\begin{figure}
\parbox[b]{6.5in}{\includegraphics[height=6.5in,angle=0]{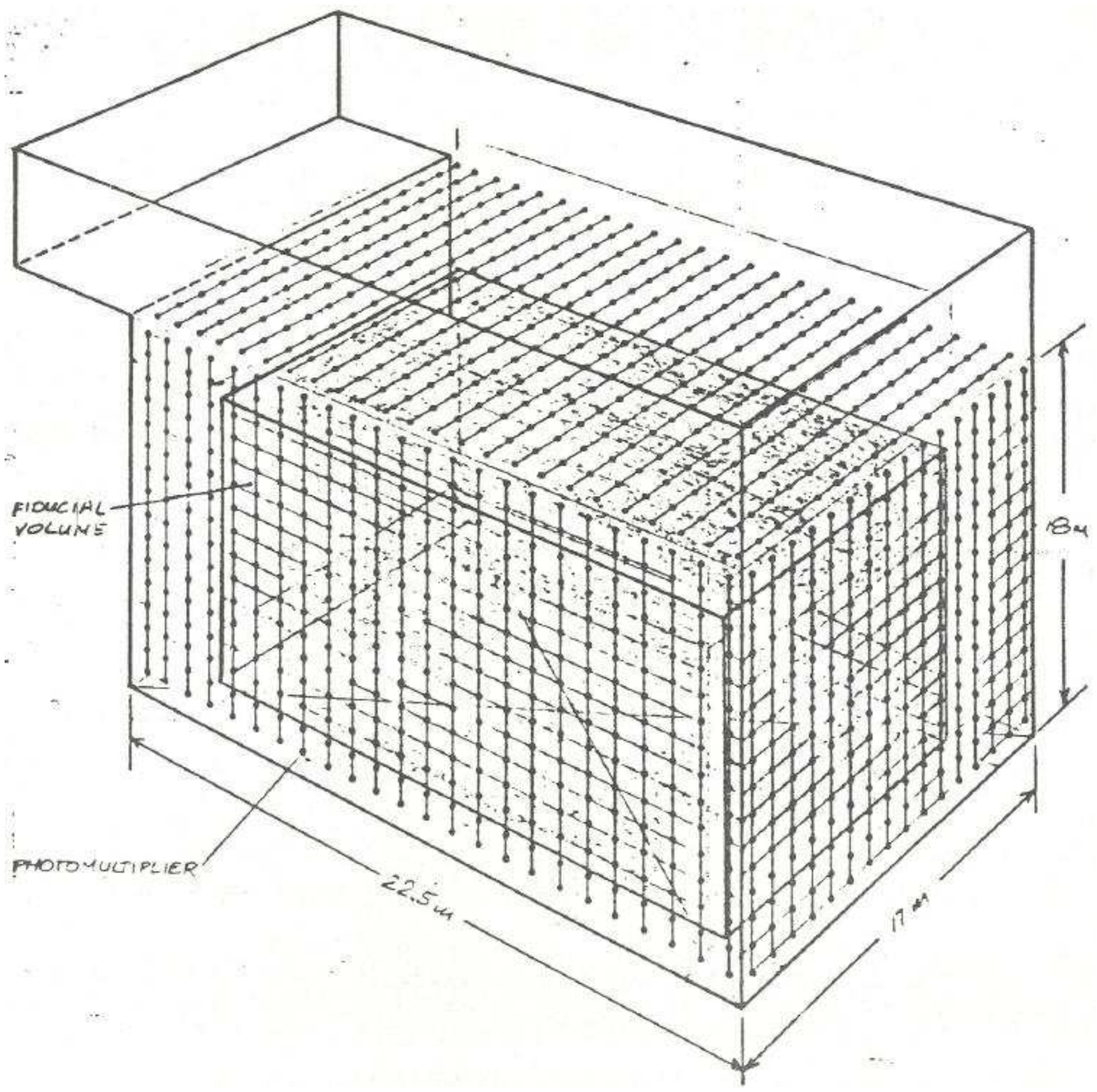}}
\caption{\label{fig:IMB-sketch-1}A sketch of the IMB detector which was used for
  much of the work described in this paper.  The sketch shows the size of the
  cavity excavated out of salt and filled with ultrapure water.  The lines and
  black dots represent the phototubes and cables used to read out light signals
  from Cherenkov light produced in the water by fast charged particles.  The
  fiducial volume two meters in from the phototube planes is indicated by the
  internal rectangular solid.  The fiducial mass is a bit over 3.3 kilotons.
}
\end{figure}
Figure \ref{fig:IMB-sketch-1} is a sketch of the detector used for this
research.  Reference \cite{SciAm} describes the search for proton decay and
includes photographs and other illustrations of the methods employed.

Most early Grand Unified theories predicted that the proton would decay
to a final state consisting of a positron and a neutral pion.  The neutral
pion would immediately decay to two photons giving a fairly clear signal.
Some variations of the theories, including those incorporating supersymmetry
predicted suppression of this decay mode but favored a decay mode containing
a charged muon and a neutral K meson.  The neutral K meson would also decay
yielding a somewhat different signature.  Some ability to distinguish the
proton decay modes was an essential part of the experiments.

An experiment searching for rare processes such as proton decay must be
very sensitive.  A sensitive detector is subject to a very large number of
detections of non-signal.  Such non-signal is termed ``background''.
A proton decay detector needs to be very large to be sensitive to the signal
but it must be well shielded to reduce the background to a level where it
does not obscure the signal.  All such low background experiments have been
located underground since cosmic rays produce a substantial contribution to
the background at the surface.  Even underground the cosmic ray rate is at best
reduced but never eliminated.  This is because the high energy muon component
of the surface cosmic rays interact primarily electromagnetically and so they
loose energy fairly slowly in traversing matter.  As a rule, the deeper the
detector the fewer cosmic ray muons one must register and reject.  One
important trade-off is that excavation costs are higher the deeper one goes.
So on a fixed budget one would have to construct a smaller detector at
larger depths.  In most cases the location of the underground laboratory
has been determined by previously existing infrastructure such as a mine or
mountain tunnel.

One form of non-signal, background, which is not attenuated by depth are
interactions from the neutrinos produced by cosmic ray interactions in the
atmosphere.  A very efficient method for producing neutrinos in the atmosphere
is by strong interaction production of pions.  The pions decay readily to a
muon and a muon neutrino.  Many of the muons also decay before they reach
the ground
to yield an electron and two additional neutrinos (one each of electron
and muon type).  Neutrinos are very
penetrating since they only interact via the weak interaction, so they are
essentially unattenuated by any amount of terrestrial shielding.

The flux
of these neutrinos can be calculated and an event rate estimated.  The
experimental challenge was to identify neutrino interactions so that they could
not be confused with the proton decay signature.  In principle it would be
easy to distinguish the two.  Protons decayed essentially at rest in the
detector, with negligible momentum whereas entering neutrinos bring momentum,
creating events with approximately equal energy and momentum.  So one can
distinguish the two classes of events by reconstructing the events and measuring
their momentum.

\section{Formulation I -- Accelerator Experiments}
\begin{figure}
\parbox{5in}{\includegraphics[width=5in,angle=2]{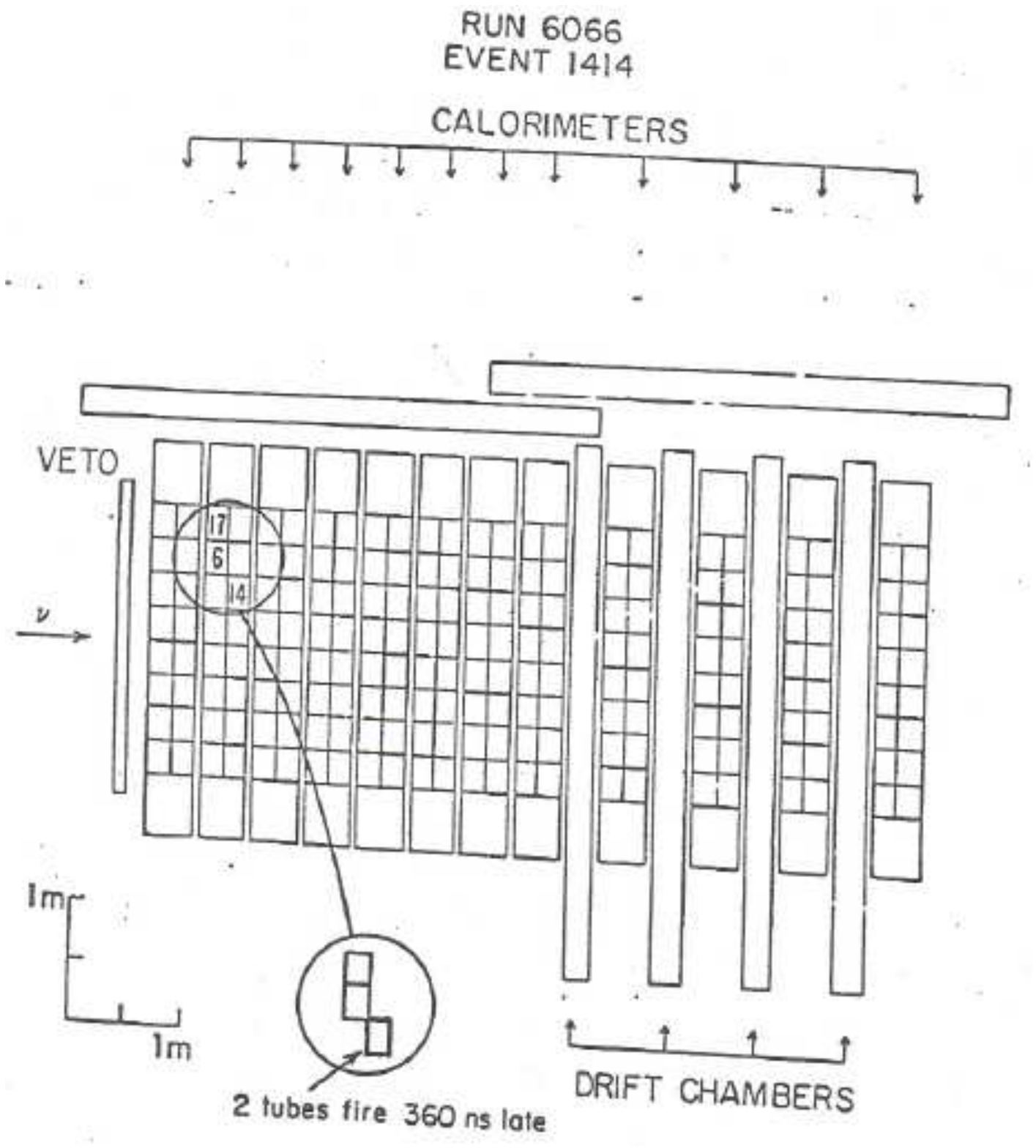}}
\caption{\label{fig:704E}An event from the Brookhaven neutrino
oscillations search.  The detector design was typical of the time.  It was
composed of a large segmented volume of liquid scintillator.  The
segmentation permitted the crude reconstruction of the tracks in the event.
In the sample event illustrated three scintillation cells have energy deposited
in them.  The recoil energy and direction, as well as a knowledge of the
neutrino beam direction, permited a reconstruction of the neutrino interaction.}
\end{figure}
The search for neutrino oscillations was begun most expeditiously by adapting
existing facilities to the project.  Once existing data had been checked, and
no evidence for oscillations found, the next step was to try to extend the
range of the searches.  One project, experiment 704 (figure \ref{fig:704E}) at
Brookhaven, utilized
a neutrino detector that had been used to establish the properties of the
weak neutral current\cite{E704}.  To extend its sensitivity to neutrino
oscillations the energy of the beam was lowered.  Lowering the neutrino beam
energy had several advantages.  It extended the sensitivity to lower
neutrino mass differences
($\Delta m^{2}$).
At the lower energies the muon neutrinos in the beam would
not interact because they had insufficient center of mass energy to produce
a muon via the charged current interaction.
\begin{figure}
\parbox{5in}{\includegraphics[width=5in,angle=270]{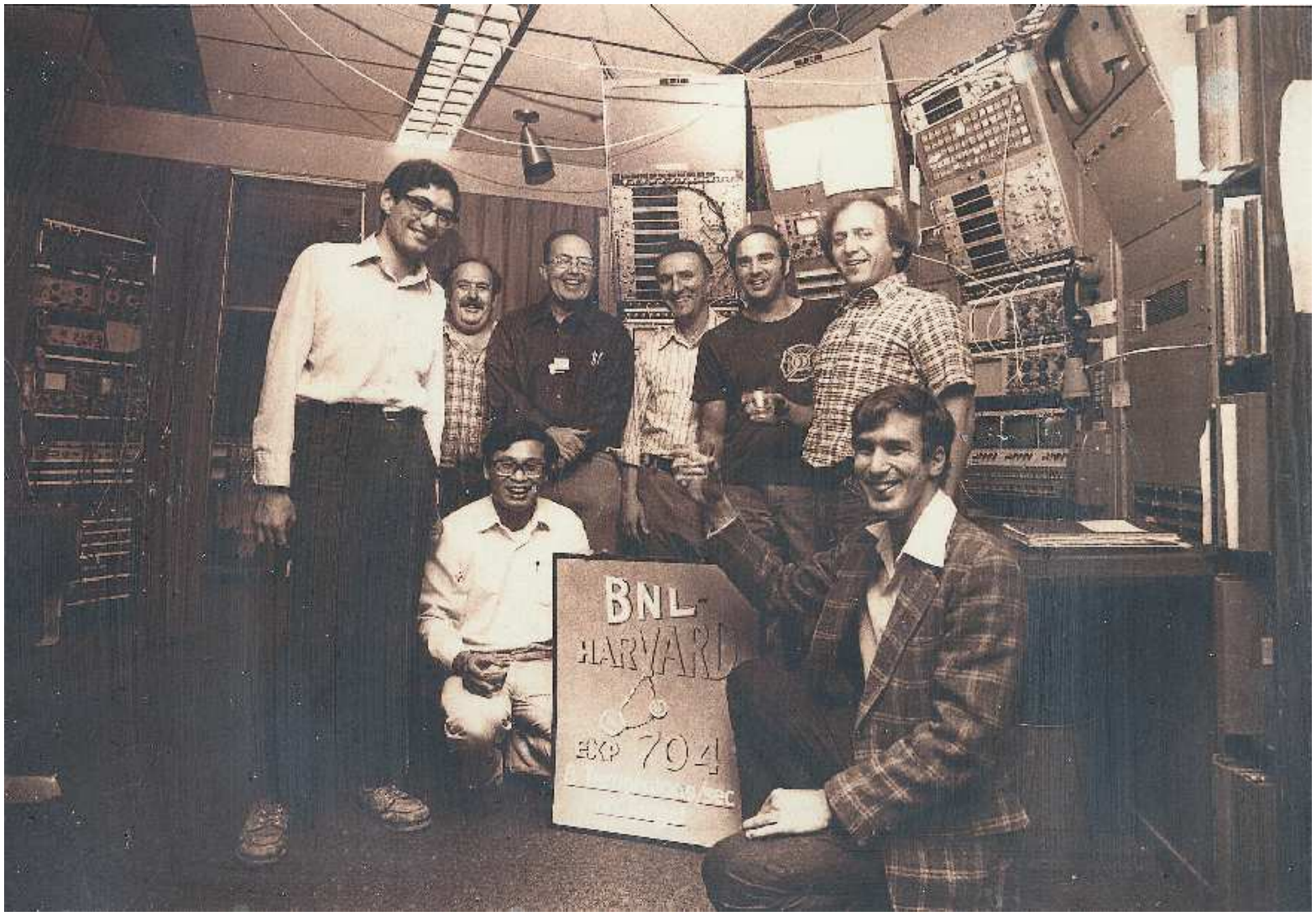}}
\caption{\label{fig:704}Members of a dedicated neutrino oscillations
experiment at Brookhaven
National Laboratory in 1977.  LoSecco is on the left and Larry Sulak is
kneeling on the right.  Bill Wang is kneeling to the right of LoSecco.
Andy Soukas is in the middle of the group of accelerator operators standing
in the back.}
\end{figure}
Interactions could only occur if the muon neutrinos transformed into electron
neutrinos.  Lowering the production energy also had the advantage that no
electron neutrinos were produced in the target since the beam was below
kaon production threshold.  Kaons are the major source of the electron neutrino
content of accelerator based neutrino beams.

The experiment failed to find evidence for neutrino oscillations but gave the
experimenters, many of whom went on to work on IMB, substantial experience
with the neutrino oscillations problem.  Among the lessons learned was a need
for a good understanding of the neutrino flux and a very good understanding of
the detector response to a potential oscillations signal. 
Figure \ref{fig:704} shows many of the participants in this early oscillations
experiment.

\section{Formulation II -- Neutrinos in Nature}
\begin{figure}
\parbox{5.5in}{\includegraphics[height=5.5in]{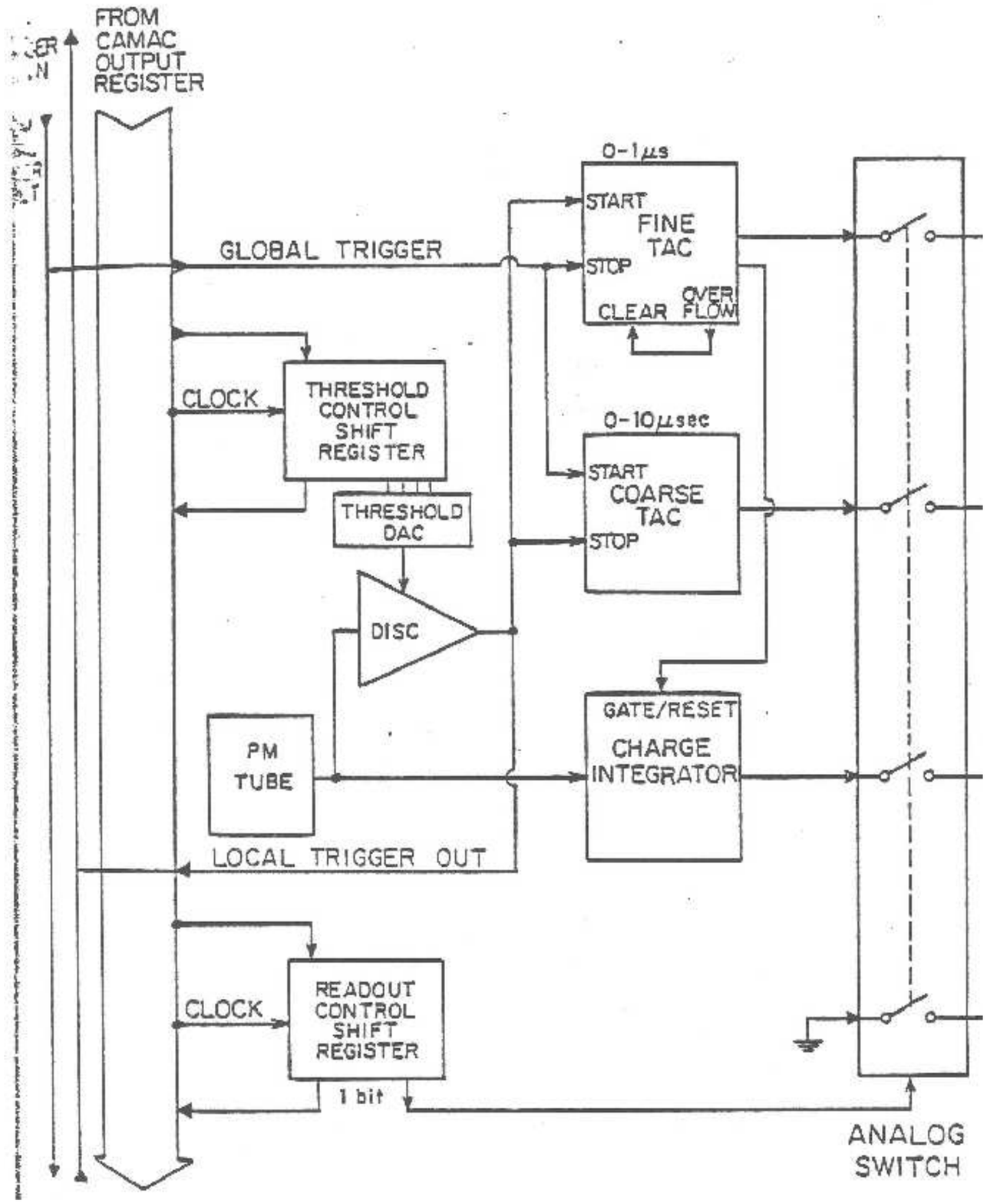}}
\caption{\label{fig:T2}The block diagram for the IMB experiment electronics.
  This provided an effective, accurate way to identify interactions leaving a
  muon in the detector.  The design utilizes the approximately 2 microsecond
  lifetime of the muon to identify it.  The electronics contained two time
  digitizers.  The top one, labeled ``Fine TAC'' had a resolution of about a
  nanosecond and produces a measurement that is used to reconstruct the position
  and direction of the event.
  The center of the figure includes the design for a coarse time scale (labeled
``Coarse TAC''), called a T2.  This
provided a 10 microsecond window, following a triggering interaction to observe
particle decays of the products of the interaction.  The rest of the electronics
on the diagram provided for energy integration, and the ability to trigger
the detector.  A muon was identified when several phototubes gave a delayed
signal following a triggered interaction.}
\end{figure}

While the flux of atmospheric neutrinos was an annoyance to the search for
proton decay, it was realized that such a signal also provided an opportunity
to extended the study of neutrino oscillations to kinematic regions of mass
differences,
$\Delta m^{2}$ well below what would be feasible at accelerators or reactors.
But to be effective the detector would have to be able to distinguish between
different kinds of neutrino interactions.  Fortunately the need to
identify the proton decay final state to distinguish different models
had similar requirements.  The IMB detector\cite{IMBprop} was designed with a
high resolution time scale to facilitate reconstruction by time of flight
and a coarse time resolution which extended out to 10 microseconds following
an event to search for a delayed signal coming from a final state muon,
$\mu \rightarrow e \nu \bar{\nu}$ (Figure \ref{fig:T2}).  This coarse time
scale, or second time scale was known as the
``T2'' scale and was to later give the name ``T2 problem'' to our discovery.
The name came about since we failed to find sufficient event candidates within
a delayed time window.  These were measured with the “T2” electronics.  A
potential problem with this part of the electronics would give a comparable
result.  Fortunately there were experimental ways to confirm the correct
operation of the electronics, by measuring stopping muons penetrating from the
surface.  So we could rule out an instrumentation problem to explain the
observations.  But the name stuck.
While massive shielded underground detectors were motivated, and funded
initially to experimentally observe predictions of Grand Unified theories,
it is noteworthy that even the earliest proposals\cite{IMBprop} clearly
indicated that such detectors would also be able to explore ``neutrino
oscillations, matter effects and supernovae''.  The groups active in the
search for proton decay, for the most part, had substantial neutrino experience
since the problems were similar in many ways.  Neutrino observation also
required massive well shielded detectors.

\begin{figure}
\includegraphics[height=4.5in,angle=0,trim= 1.5in 7.65in 1.5in 0in,scale=1.,clip=true]{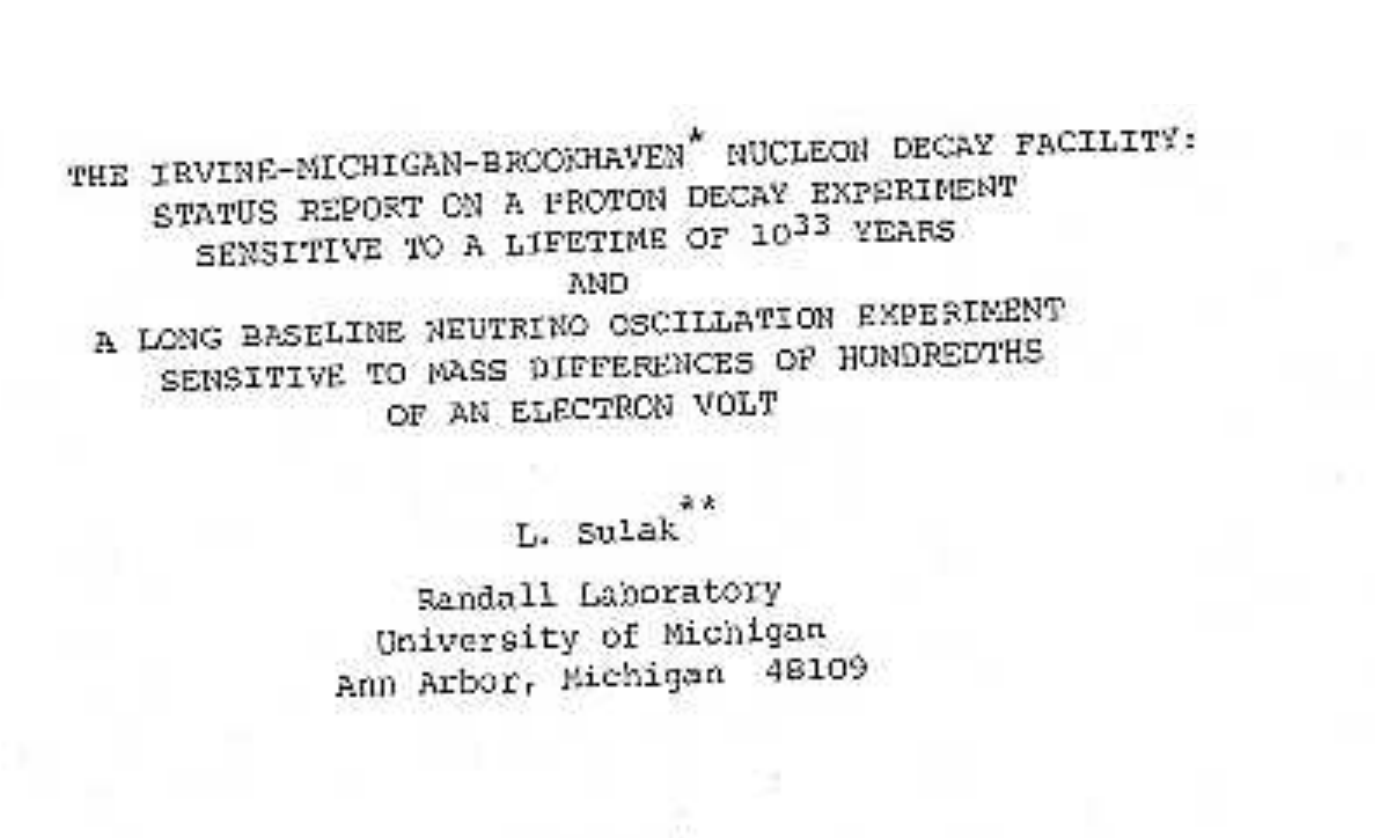}
\caption{\label{fig:FWOGU}The title page for Sulak's talk about the potential
to observe
the oscillations of atmospheric neutrinos presented at the 1980 First Workshop
on Grand Unification (FWOGU) in New Hampshire.}
\end{figure}

While the IMB proposal\cite{IMBprop} had mentioned neutrino oscillations the
details were not filled in until spring of 1980.  As part of the graduate
school requirements at Harvard University, students were expected to prepare
a project on a topic related to their thesis research and to present it as an
oral exam.  Bruce Cortez, a student of Larry Sulak, chose atmospheric
neutrino oscillations as his qualifying orals topic\cite{BGC}.  His work was
fairly complete.  It included details of the neutrino flight distance as a
function of direction.  Upward going neutrinos travel about 13,000 km but
those going down only travel about 15 km from their point of production in the
atmosphere.  The neutrino direction is determined from the direction of the
momentum of the reconstructed neutrino interaction.  
Though, in principle one got all
of the distances in between the transition from up to down distance scales
occurs very rapidly over a fairly small part of the total solid angle near
the horizon so,
in essence one was dealing with an approximately 2 distance neutrino
experiment.  The correlation of neutrino direction, with the direction
reconstructed from the final states was studied.  For most neutrino events
the reconstructed direction of the outgoing muon or electron provides a
reasonable estimate of the neutrino direction (and hence path length) but this
concordance tends to be less reliable at lower energies.  Fortunately the
problem does not prevent telling up from down.  The work showed that for about
two orders of magnitude $\Delta m^{2}$, from below 10$^{-4}$ eV$^{2}$ to about
10$^{-2}$ eV$^{2}$ one would have a very clear difference in the electron to
muon ratio measured for upward events compared to the observed value for
downward events.  The downward events provided a short range sample to which
the long range upward going events could be compared.  Figures prepared for
this work also showed a substantial distortion in the neutrino spectra
if $\Delta m^{2}$ were in a range just below or just above the one where
the effect would be maximal.
\begin{figure}
\includegraphics[angle=270,width=6.5in]{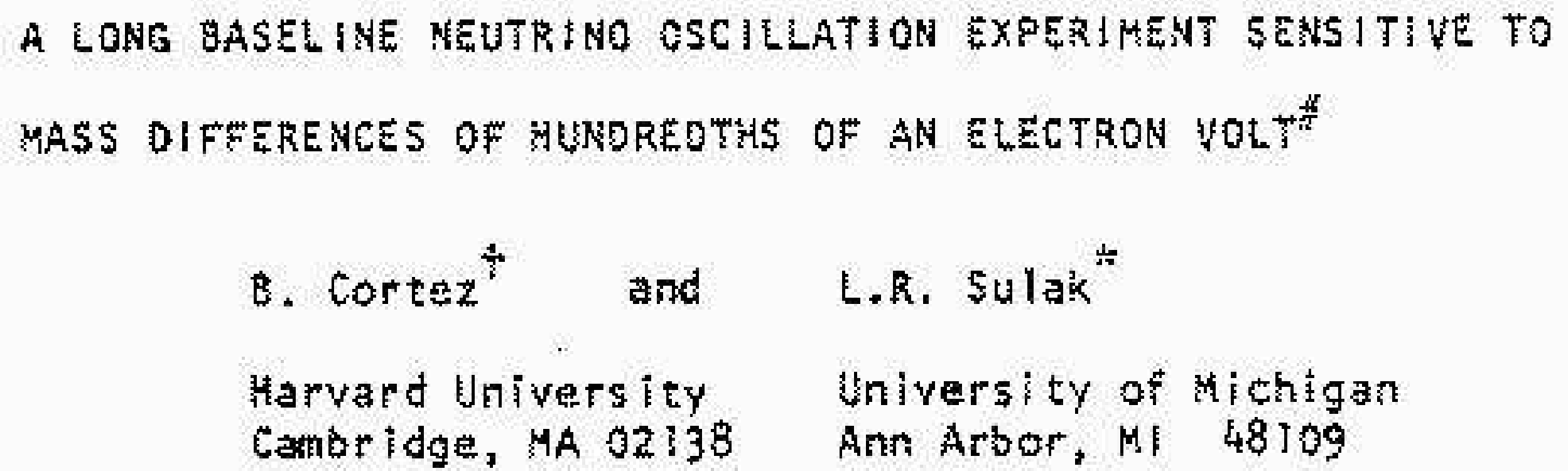}
\caption{\label{fig:Erice}The title page for Sulak's talk about the
potential to observe
the oscillations of atmospheric neutrinos presented at the 1980 EPS meeting
in Erice, Sicily.}
\end{figure}
The Cortez oral work was documented in a number of conference
talks\cite{Erice80,FWOGU-LRS}.  Figures \ref{fig:Erice} and \ref{fig:FWOGU}
are the title pages from some of these talks.
\noindent
\begin{figure}
\includegraphics[height=7.5in,angle=0]{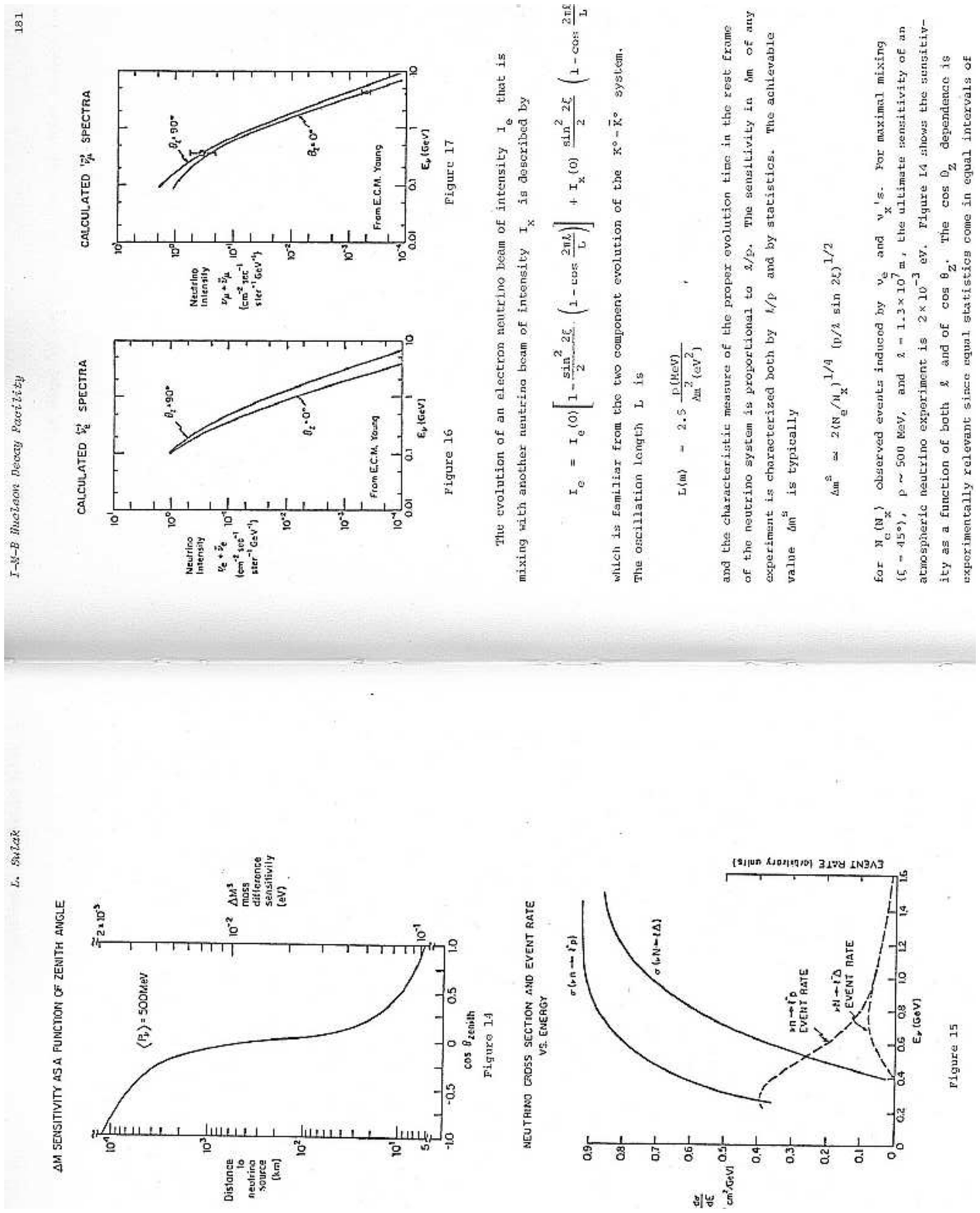}
\caption{\label{fig:ANOa}Figures taken from the FWOGU and Erice talks
illustrating the method
to observe atmospheric neutrino oscillations with a detector that was already
under construction to observe proton decay.  The neutrino propagation
distances and the neutrino energy spectrum are shown here.}
\end{figure}
\begin{figure}
\includegraphics[height=7.5in,angle=0]{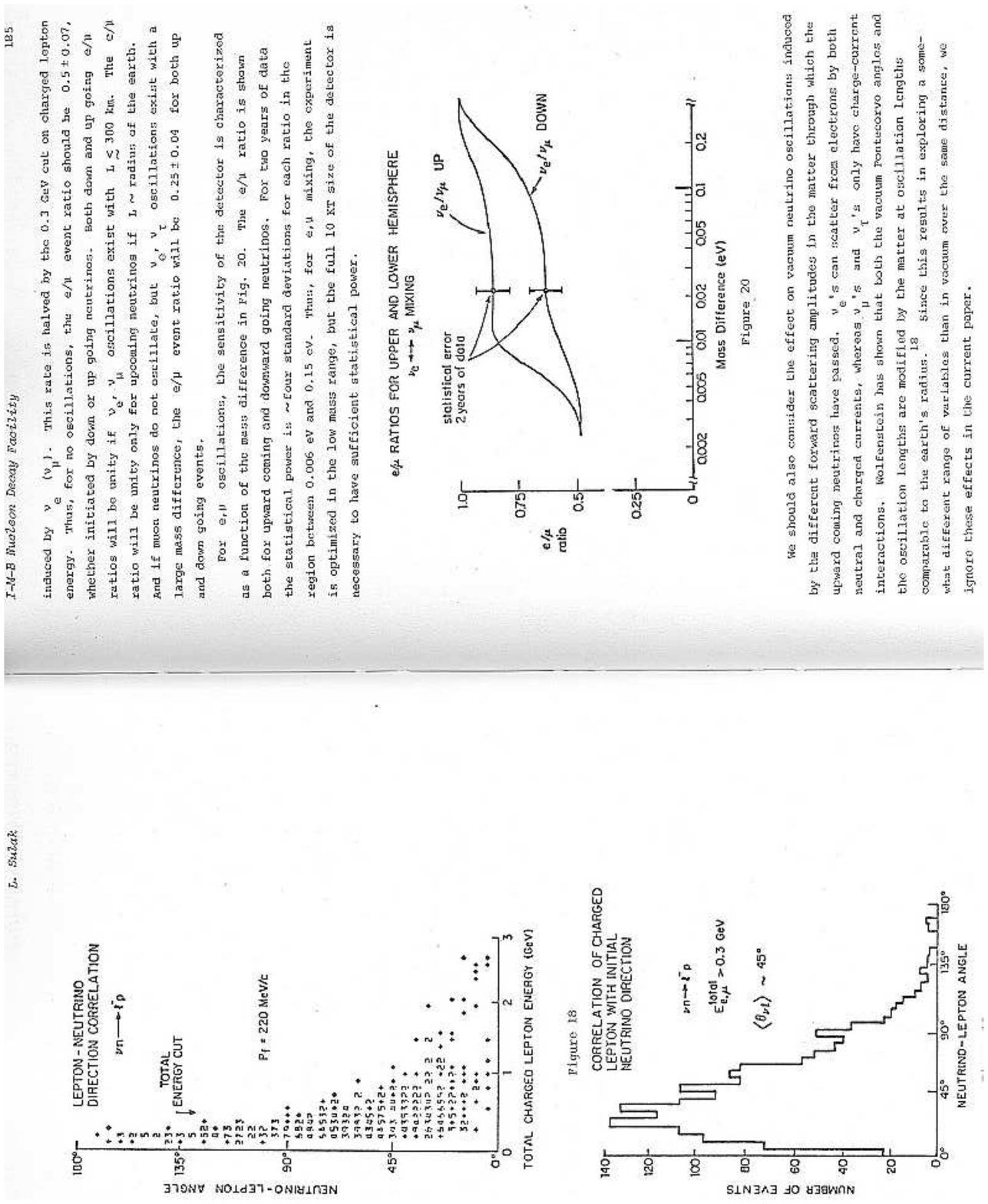}
\caption{\label{fig:ANOb}Figures taken from the FWOGU and Erice talks
illustrating the method
to observe atmospheric neutrino oscillations.  The neutrino energy and direction
resolution are considered here.}
\end{figure}
\begin{figure}
\includegraphics[height=6in,angle=270]{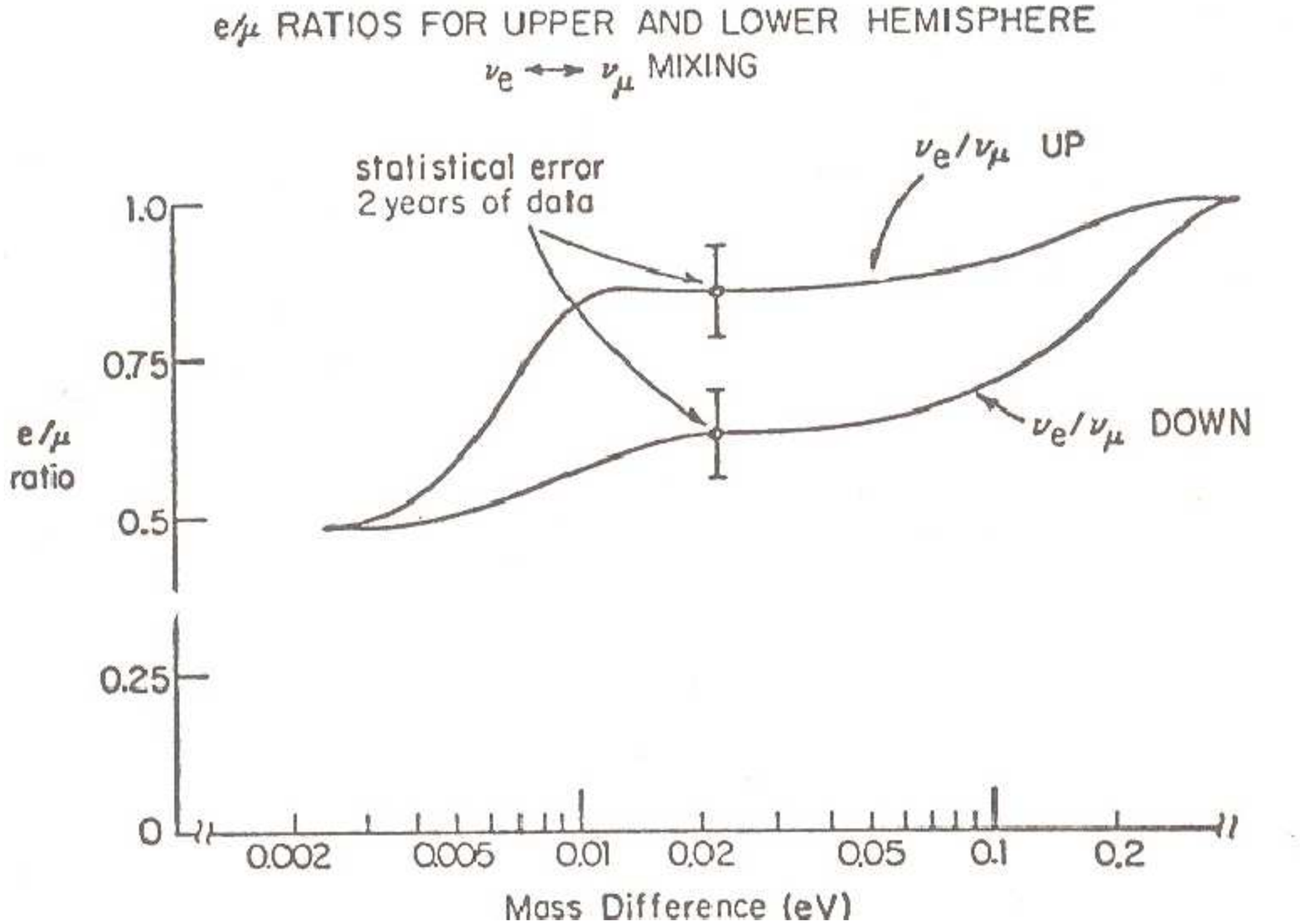}
\caption{\label{fig:ANO2}The expected ratio of electron neutrino
interactions to muon neutrino
interactions for those neutrinos going upward and those going downward plotted
as a function of the (positive) neutrino mass difference.  The difference
between these two samples is sensitive to neutrino oscillations over a
substantial range of mass differences.}
\end{figure}
Figures \ref{fig:ANOa} and \ref{fig:ANOb} summarize most of the information
needed to study
atmospheric neutrino oscillations including the neutrino flux and cross
sections, the variable distances and resolution issues.  Figure \ref{fig:ANO2}
emphasizes the primary observable would be a difference in the electron
to muon neutrino interaction rate as a function of neutrino direction.

So the atmospheric neutrino sample collected in the detector would include
neutrinos with both short and long travel distances.  Comparison of these
samples would provide evidence for neutrino oscillations.  The short
ones would provide a control sample of unmodified neutrinos that could be
compared with the longer flight ones which could have oscillated over the
extra flight time.

\section{Preparation}
\begin{figure}
\includegraphics[width=5in,angle=270]{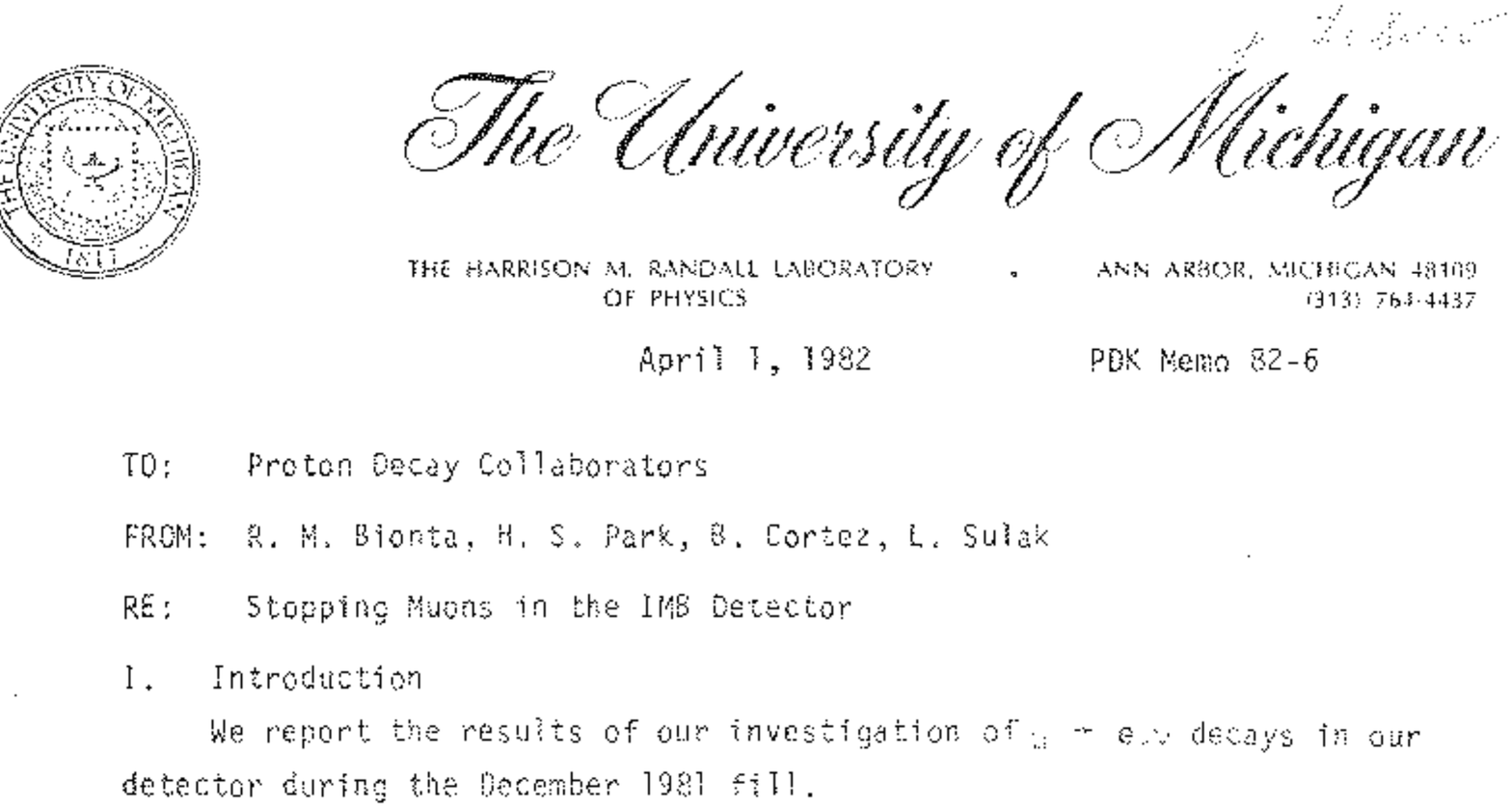}
\caption{\label{fig:T2memo}Heading from a 1982 internal report on calibrating
the detector
response to muon decays.  This work was done well before the detector was
completed}
\end{figure}

Detector construction took up much of the period of 1980-1981.  A first attempt
to fill the detector was made in December 1981.  Some flaws in design were
discovered at that time but even the partial fill yielded important data
on the detector performance.  A memo\cite{Bionta1982} (figure \ref{fig:T2memo})
from early 1982
demonstrated the detector response to stopping comic ray muons even though only
1/3 of the detector had been filled.  The particle identification system, the
``T2'' scale was validated.  Muons from the surface stopped in the partially
filled detector gave the expected delayed signal about 2 microseconds after
they had stopped.  Timing and energy distributions met expectations.  The muons
triggered the detector when they entered it.  The subsequent muon decay
populated the delayed time scale electronics.
A substantial effort was made during the initial start-up period to understand
the detector.  One had to demonstrate that the device saw what was expected
to be there, atmospheric neutrinos and cosmic ray muons, before one could
believe that it was also capable of observing proton decay.

Since the
atmospheric neutrinos were expected to be the only serious background to proton
decay several efforts were made to control systematic errors associated with
them.  The atmospheric neutrino response in the detector was modeled using
real neutrino interactions.  We had access to the large sample of accelerator
neutrino interactions acquired at CERN in the heavy liquid bubble chamber
``Gargamelle''.  These interactions were primarily on bromine, a slightly
heavier nucleus than the oxygen found in our water.  But we needed a sample
of neutrino interactions on nuclei since these would include absorption,
rescattering and Fermi motion effects caused by the other nucleons in the
nucleus.  Subsequent to the work with the ``Gargamelle'' events we also
studied events on neon from Brookhaven and on deuterium from Argonne.
Several neutrino interaction models were also prepared to facilitate comparison
and gauge systematic error.

The experiment was fortunate in that it had access to a large convenient
sample of stopping muons to calibrate the detector response to muon decay.
\begin{figure}
\parbox[t]{3.8in}{
\raisebox{2in}[1in][1in]{\includegraphics[width=4.25in,angle=270]
{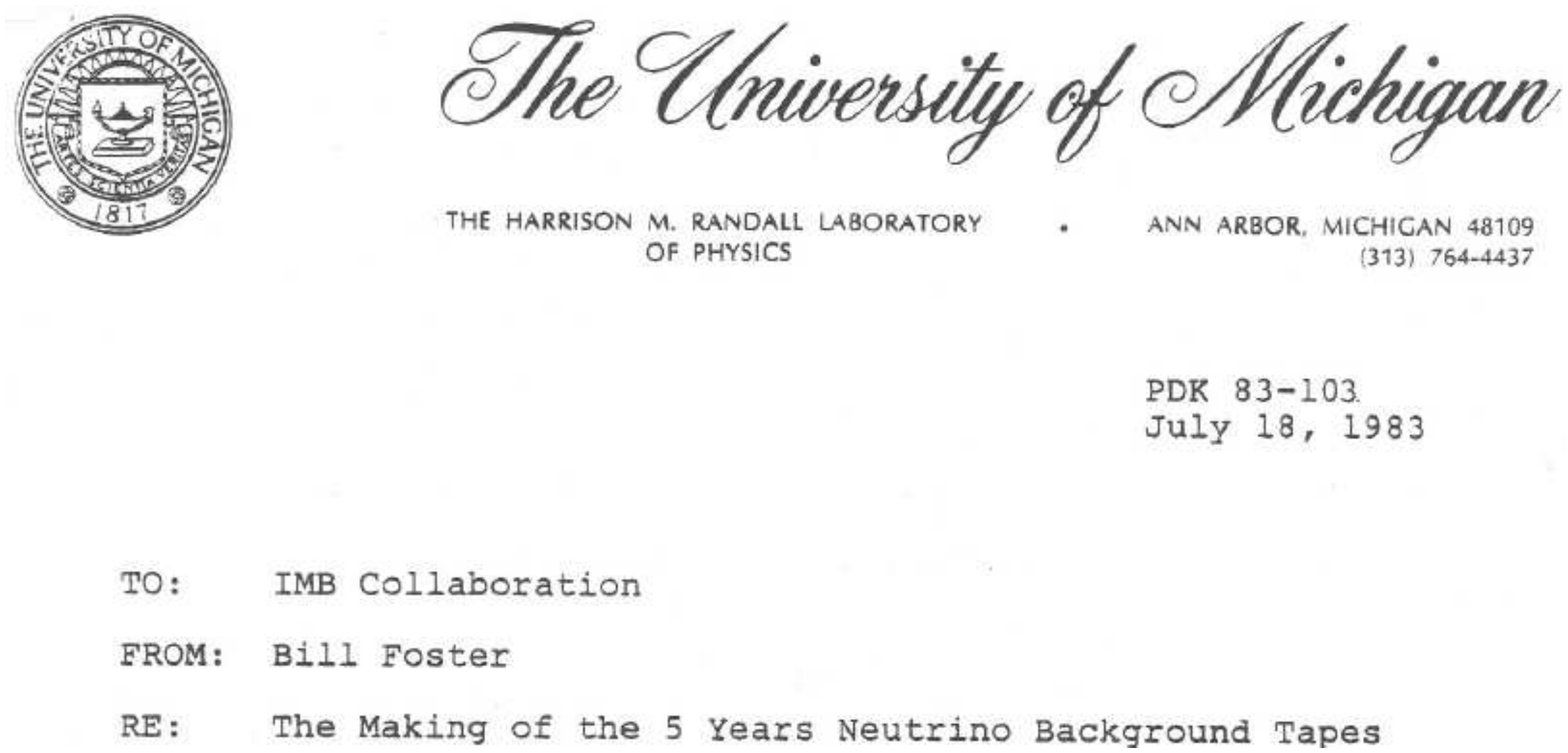}}\\
\raisebox{1.9in}[1in][1in]{\includegraphics[width=4.25in,angle=270]
{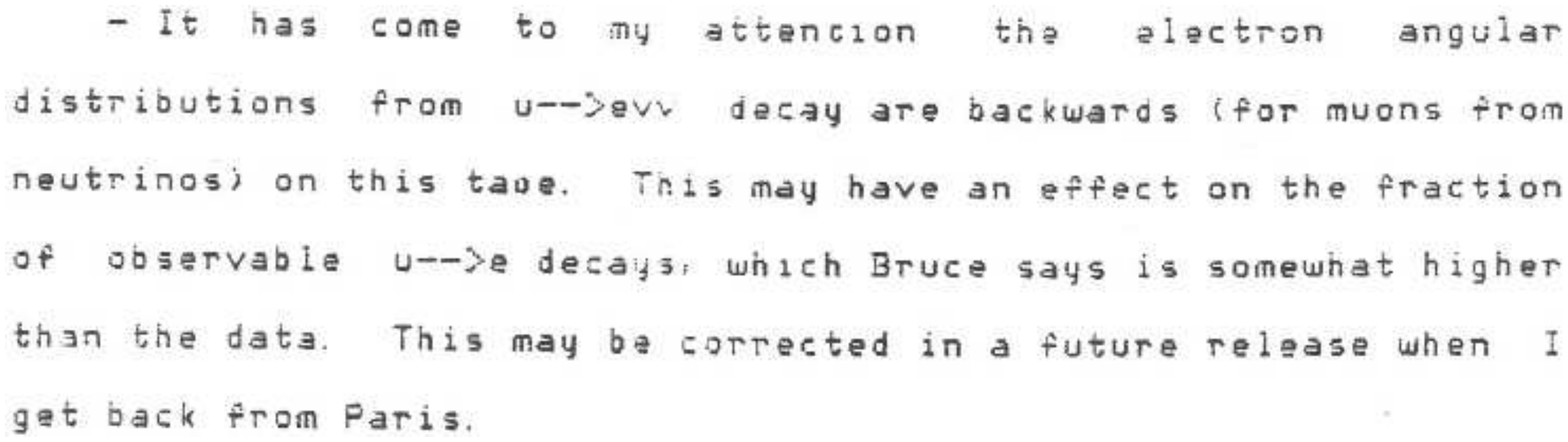}}
}
\caption{\label{fig:Disc}Portions of Bill Foster's memo of July 1983, which
describes the way in which
neutrino interactions were simulated in the detector.  At the end of this
memo he notes a {\em discrepancy} between the observed number of muon decays
and the number expected based on this simulation.}
\end{figure}
In a memo of summer 1983\cite{Foster1983} (figure \ref{fig:Disc}) Bill Foster
described the way in
which ``Gargamelle'' events were converted to IMB events and were simulated to
understand the background.  That note has a very interesting concluding
paragraph.

``It has come to my attention that the electron angular distribution from
$\mu \rightarrow e \nu \nu$ decay are backwards (for muons from neutrinos)
on this tape.  This may have an effect on the fraction of observable
$\mu \rightarrow e$ decays, which Bruce says is somewhat higher than the data.
This may be corrected in a future release when I get back from Paris.''

The simulation had more muon decay events than had been observed in the
detector.  This became know as the “T2 problem”.  Something needed to be
understood.  The search was on to find some systematic error that could account
for the discrepancy.  As the quote indicates, the muon polarization was
considered a possible candidate for the cause.  The physical significance of
this remark is that the spin direction of muons coming from neutrino
interactions is opposite that of muons coming from pion decay.  The calibration
response based on stopped muon decays could be slightly different than the
detector response to muons formed by neutrino interactions.

\section{Observation}
The September 1983 Harvard PhD. theses of Bruce Cortez and Bill
Foster\cite{HarvPhd}
contained the first physics to emerge from the class of experiments
initiated to discover proton decay.  Their data sample consisted of 112
contained events collected over a period of 130 days.  They had searched
for proton decay into the decay modes, lepton $K^{0}$ and $e^{+} \pi^{0}$.
No evidence for proton decay had been found.  The data sample included 25
events which had a muon decay.  This was 22$\pm$4\% of the sample when 33\%
had been expected.

This muon deficit, which was 2.5 standard deviations too low, was the net
result of the discrepancy mentioned at the end of the previous section.

Followup of this ``T2 problem'' came with work by Eric Shumard, summarized in
his 1984 PhD. thesis\cite{Shumard} for the University of Michigan.  A
major portion of
Shumard's thesis was devoted to extensive study of the IMB detector's response
to muon decay.  He did a very careful job of measuring and modeling the muon
identification process.  He included all effects such a muon polarization,
absorption, reflections of light and after-pulsing of phototubes.  The thesis
was based on 148 contained events collected over 202 days.  The sample included
39 events with muon decays, which was 26.4$\pm$3.6\% but 35\% of the sample,
52 events, were expected.
Shumard had succeeded in improving the responsiveness of the detector to muon
decays, and hence the ability to recognize muon neutrino interactions.
But in spite of this effort the observations were still about 2.4 standard
deviations below expectations.

As was typical at the time, the topic of Shumard's PhD. thesis was proton
decay.  He also reported no evidence for this process.

The IMB-1 detector ran for 417 days of live time before undergoing
modifications to what eventually became IMB-3.  The raw data from IMB was
analyzed twice, by two independent programs to increase the detection
efficiency and to guard against potential systematic bias.  One of these
streams was based at Caltech and Irvine and was known as the ``West coast''
analysis.  The October, 1985
Caltech PhD. thesis of Geoff Blewitt\cite{Blewitt} included 326 contained
events, from only
the ``West coast'' version of the reconstruction streams.  He reported a
muon decay rate that was 2.8 standard deviations too low.

The full IMB-1 data sample consisted of a merger of both analyses.  It had
401 contained events.  Of these 104 had a muon decay.  This 26$\pm$2\% observed
muon decay rate was 3.5 standard deviations below the expected rate of
34$\pm$1\%.

The evidence for an atmospheric muon neutrino deficit emerged gradually as
data accumulated and consistency checks were made.  At the IMB depth there was a
reasonable source of stopping surface muons that provided a valuable resource
to ensure we understood the detector response.  Statistical errors drop with
time as more data is accumulated but experiments can be limited by systematic
effects which, in general, are not improved by more data.  Systematic effects
can be controlled by comparison with a known
signal, such as the pure sample of muons from pion decay in the atmosphere that
penetrate the Earth and stop in the detector.  Small
details in experimental design, such as the delayed time coincidence of the IMB
electronics can make a big difference in the control of these systematic
effects.
\begin{figure}
\includegraphics[width=3.25in,angle=270]{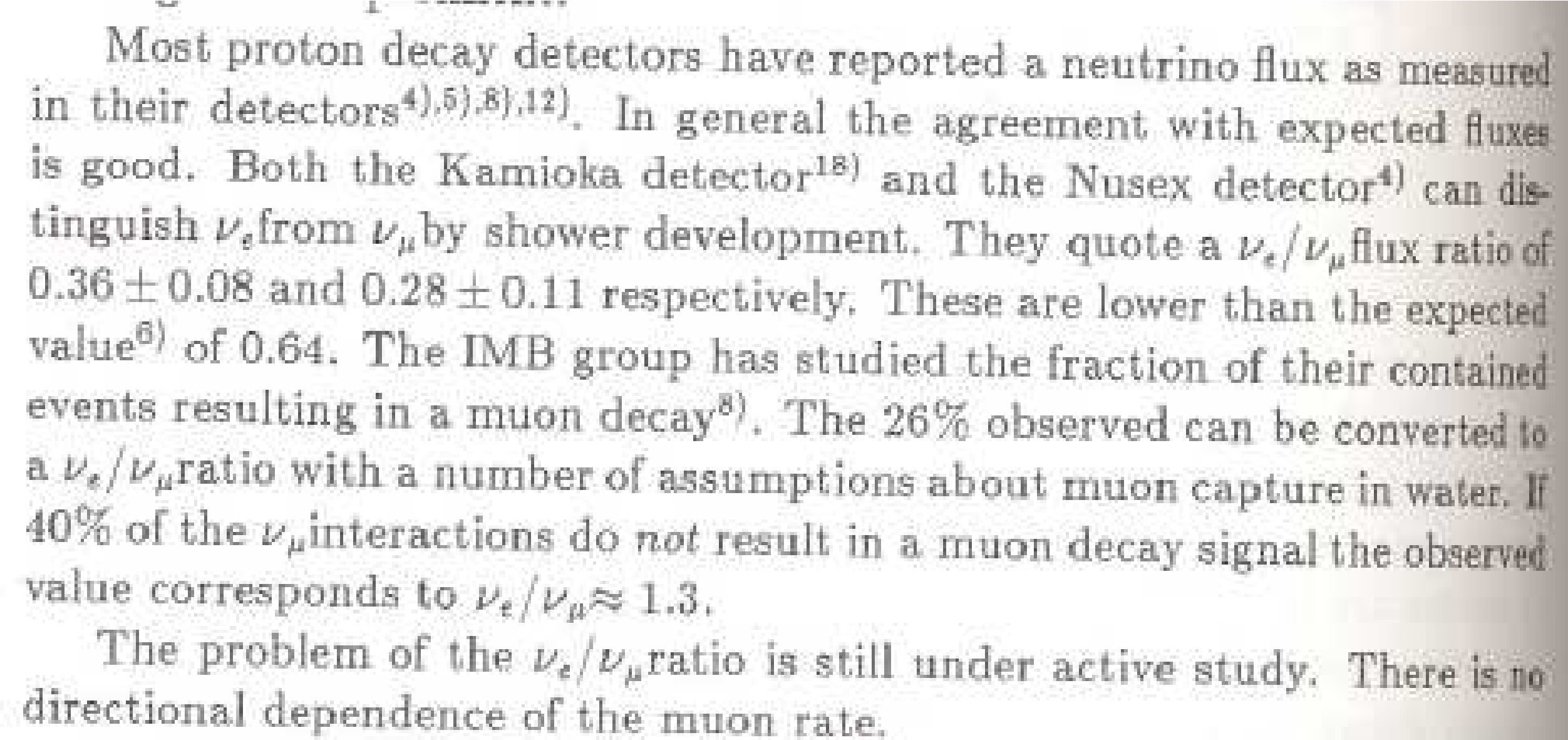}
\caption{\label{fig:LL}A portion of the proceedings of the 1986 Lake Louise
meeting in
which I point out a discrepancy between the IMB muon observations and
expectations and the reports of two other experiments.}
\end{figure}

In February of 1986 I was invited to give a talk at the Lake Louise
meeting\cite{LLouise} (a series of winter institutes held in the
Canadian Rockies )
summarizing the status of the search for proton decay.  As part of the talk
I reviewed the atmospheric neutrino observations (figure \ref{fig:LL}).  I
mentioned the IMB muon
discrepancy of 26\%.  ``If 40\% of the $\nu_{\mu}$ interactions do not result
in a muon decay signal the observed value corresponds to $\nu_{e}/\nu_{\mu}$
of 1.3''.   The expected value for such a ratio was 0.64.  Two other
experiments had values for this observable which I reported.  The Nusex
experiment, an iron calorimeter in the Mont Blanc tunnel\cite{Nusex} had
reported a value of 0.28$\pm$0.11.  The Kamioka experiment, using a novel
method to distinguish showering from nonshowering events in water detectors had
reported\cite{KoshAspen} a value of 0.36$\pm$0.08.  I had no explanation for
the apparent discrepancy between the IMB observations and the two other
experiments.

\begin{figure}
\includegraphics[height=4in,angle=270]{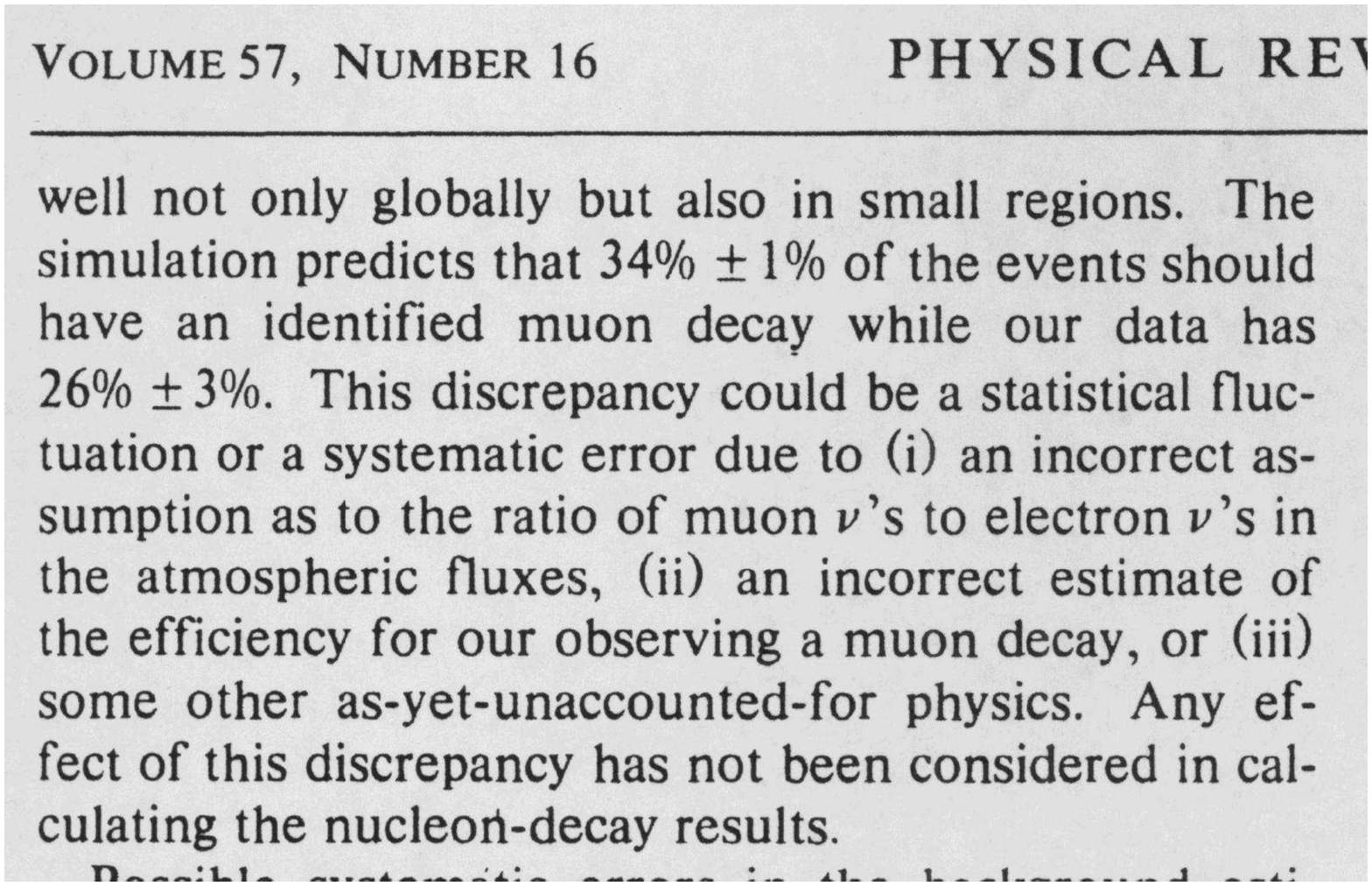}
\caption{\label{fig:NUtext}Excerpt from the 1986 IMB neutrino paper
describing the atmospheric
neutrino anomaly and some potential explanations.  The variety of explanations
represented the varied opinions of the multiple authors.}
\end{figure}

As part of the effort to understand the background to proton decay there were
extensive efforts to understand the details of neutrino interactions and their
final states.  The use of actual bubble chamber neutrino interactions
in early simulations was a way to avoid confronting the problem but detailed
models were produced and compared with observations.  One of the first to
construct a model and compare it with the data was Todd Haines of
Irvine\cite{Haines}.  The agreement was quite good, except for the observed
rate of muon decays.
\begin{figure}
\includegraphics[height=6.5in,angle=270]{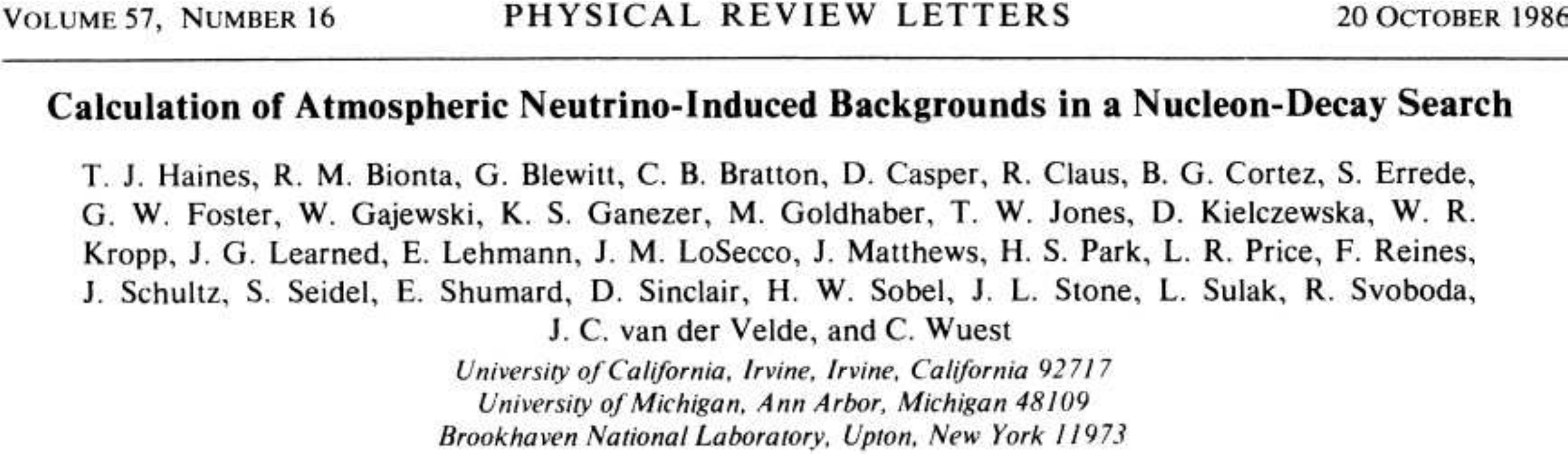}
\caption{\label{fig:NUAN}The title and author list from the IMB journal article which noted
the atmospheric neutrino anomaly.}
\end{figure}
This work on modeling the neutrino background was
published\cite{Anom} in 1986 (figure \ref{fig:NUAN}).  The paper noted the muon
rate discrepancy.
The title of the paper ``Calculation of Atmospheric Neutrino-Induced
Backgrounds in a Nucleon-Decay Search'' was appropriate for a paper comparing
observations with estimates.  The paper did not provide an explanation
(figure \ref{fig:NUtext}).
``This discrepancy could be a statistical fluctuation or a systematic error
due to (i) an incorrect assumption as to the ratio of muon $\nu$'s to electron
$\nu$'s in the atmospheric fluxes, (ii) an incorrect estimate of the efficiency
for our observing a muon decay, or (iii) some other as-yet-unaccounted-for
physics.''  The diversity of interpretations reflected the diverse opinions
of the authors.  In reality, the first two possible hypotheses could
at best reduce the statistical significance of the result.  Any uncertainty
in the flux
or the muon decay rate could not correct for the apparent 40\% reduction
in the muon neutrino interaction rate the observations suggested.  The large
scale of the anomaly was reflected in my earlier Lake Louise quote above.
After correcting for inefficiencies the muon neutrino data was almost a factor
of 2 off from expectations.

It is noteworthy that most collaborations, including IMB can be very
conservative.  As can be clearly seen in many documents leading up to this
period, such as the PhD. theses quoted, most people hoped that the effect
\begin{figure}
\includegraphics[height=5.75in,angle=90]{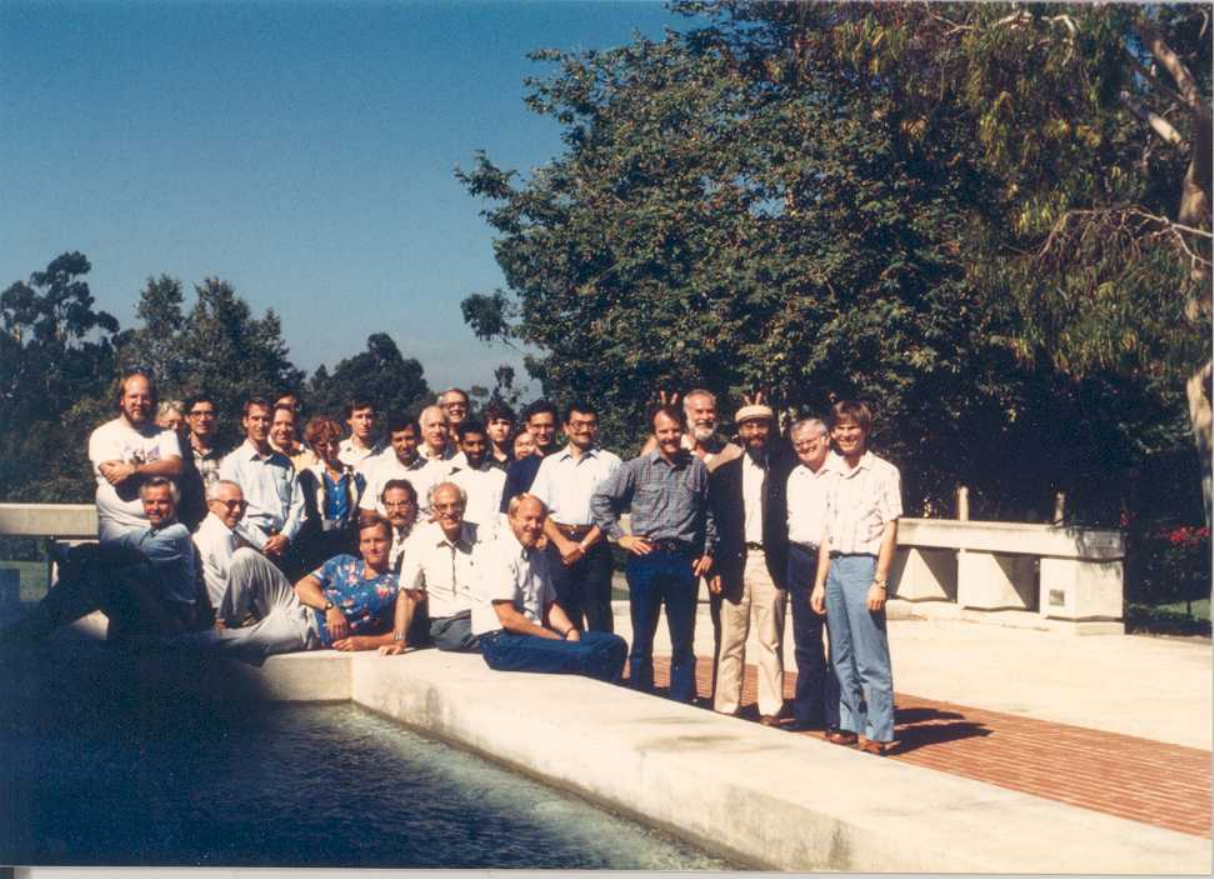}
\caption{A group photo of the IMB collaboration in 1987.  The photograph
was taken at a meeting in Irvine in which the observation of neutrinos
from supernova 1987a was celebrated.}
\end{figure}
would just go away since it constituted an uncertainty to the background to
proton decay.  In fact the muon decay deficiency was not mentioned in
early drafts of the 1986 article.  It was added, at my insistence, since the
topic of the paper, comparing neutrino observations with expectations, seemed
appropriate.

While people have voiced criticism of the wording used in this\cite{Anom}
paper, because of the multiple hypotheses provided.  The multiple hypotheses
was a compromise among the authors that permitted a significant
effect to be reported to a larger audience.  This journal
publication alerted the scientific community to an important discrepancy of
the muon neutrino rate when both Nusex\cite{Nusex}
and Kamiokande\cite{KoshAspen} had reported no such problems.
\begin{figure}
\includegraphics[height=6in,angle=0]{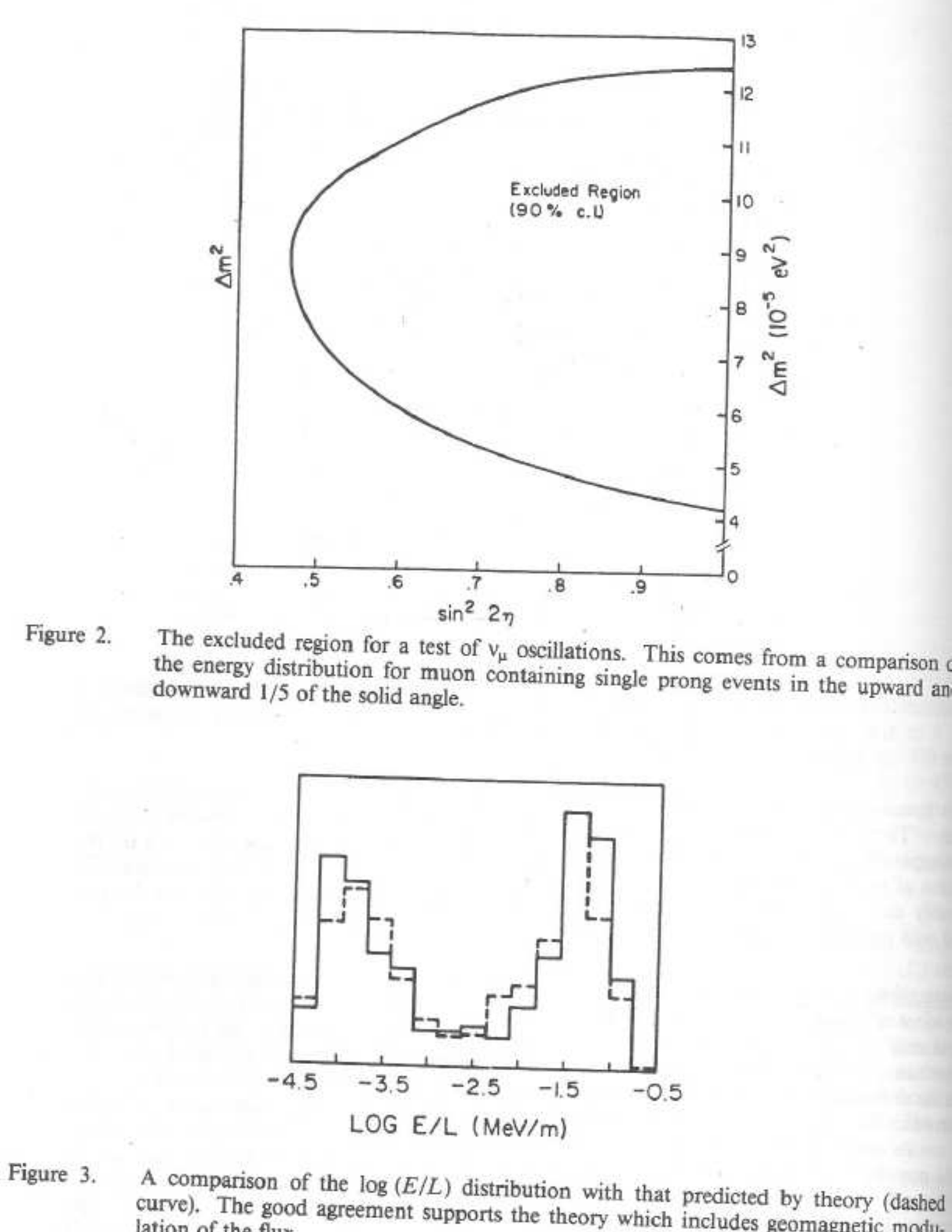}
\caption{\label{fig:SD}Illustrations taken from the proceeding of the cosmic
ray conference in San Diego in 1985.  The upper part of the figure shows a
neutrino oscillations exclusion region calculated based on the absence of
spectral distortion in comparing the upward going to downward going neutrino
events with a muon decay signature.
The lower part of the figure compares the expected and observed E/L
distribution for neutrino interaction in IMB-1.  A problem in the -2.23 bin
was noted in the text.}
\end{figure}

The error reported on the observed muon decay rate in the published
letter\cite{Anom} was $\pm$3\% rather than the $\pm$2\% mentioned earlier.
The smaller value, calculated using binomial statistics is correct because
binomial statistics ensures
that the error on the fraction of events with a muon decay is exactly the same
as the error on the fraction of events without a muon decay.  The error
can be easily calculated from the numbers in the paper.

\section{Interpretation}
\begin{figure}
\includegraphics[height=6in,angle=0]{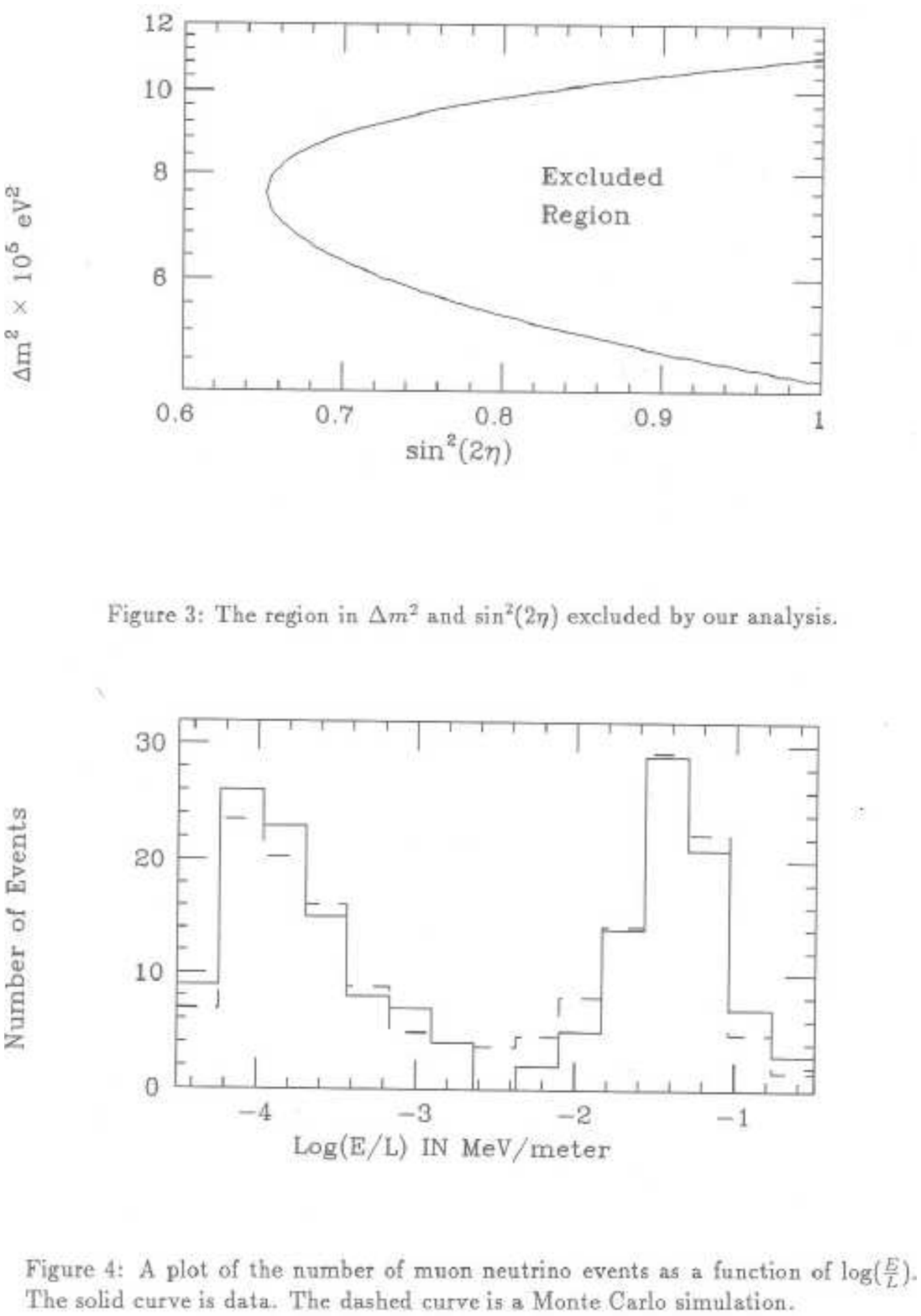}
\caption{\label{fig:BSV}IMB-3 data similar to figure \ref{fig:SD}.
This figure was shown
at a cosmology meeting in the Baksan Valley of the Soviet Union in 1991.
The lower distribution is similar to that of figure \ref{fig:SD} but this figure
includes only events with an identified muon.  In figure \ref{fig:SD} it
included all single track contained events from the earlier IMB-1 data sample.}
\end{figure}
Interpretation of the observations was difficult.  Except for the
deficiency of the muon decay rate, distributions did not look like those
expected from neutrino oscillations as studied in the early 1980 work by Cortez
and Sulak\cite{Erice80,FWOGU-LRS}.
Two component neutrino oscillations can be described by the equation
\[
P(\nu_{\mu}\rightarrow\nu_{\mu})=
1 - \sin^2({2\eta}) \sin^2({1.27 \Delta m^{2} \frac{L}{E}})
\]
  Where $\sin^2({2\eta})$ and $\Delta m^{2}$ are constants of nature that
  determine the magnitude and time scale of the effect.  A sign of neutrino
  oscillations is given by the $\frac{L}{E}$ dependence, where $L$ is the
  distance the neutrino has traveled and $E$ is the neutrino energy. 
  
There was no up down asymmetry which might be expected because of the different
distances traveled by the downward and upward going neutrinos.
There was no distortion of the energy
spectrum.  The event rate was as expected.  401 events had been observed when
402 were expected.

Comparison of the neutrino energy divided by neutrino
flight path (E/L) showed only minor differences\cite{LOE} between data and
simulation (figure \ref{fig:SD}).  The article\cite{LOE} did comment on the poor
fit of the data to expectations in figure \ref{fig:SD} by noting that most of
the $\chi^{2}$ came from the ``E/L = 5.8 $\times$ 10$^{-3}$ MeV/meter.''

This absence of unusual distributions was also found in the IMB-3 data
sample\cite{LOE3} (figure \ref{fig:BSV}) taken many years later.  The IMB-3
sample was independent.  The IMB-3 detector had four times the light collection
of IMB-1 and utilized several pattern based particle identification methods
to find muon neutrino interactions independently of the muon decay signal.
The text of \cite{LOE3} also noted a poor fit for the bin corresponding to
``$\Delta m^{2}$ of 4$\times$10$^{-3}$ eV$^{2}$''.

The normality of distributions was used to publish limits on neutrino
decay\cite{NuDK}, matter effects\cite{Matter} and neutrino
oscillations\cite{NuOsc} in the range $\Delta m^{2}<10^{-4}$ eV$^{2}$.

\section{Consternation}
\begin{figure}
\includegraphics[height=3in,angle=0]{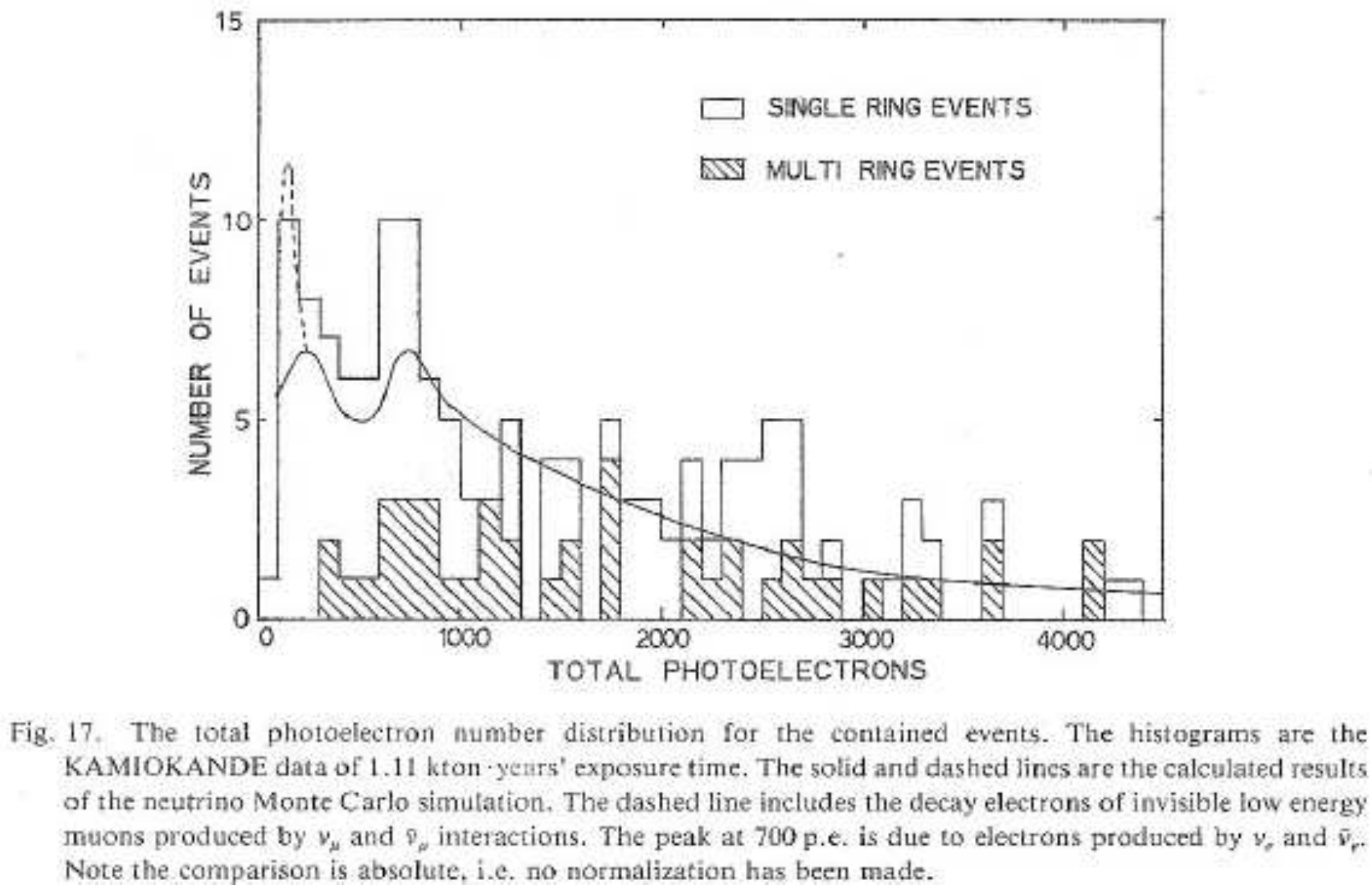}
\caption{\label{fig:NK1}The atmospheric neutrino spectrum observed with Kamiokande I.
The histogram data and the curve simulation are in good agreement.}
\end{figure}
\begin{figure}
\includegraphics[height=3in,angle=0]{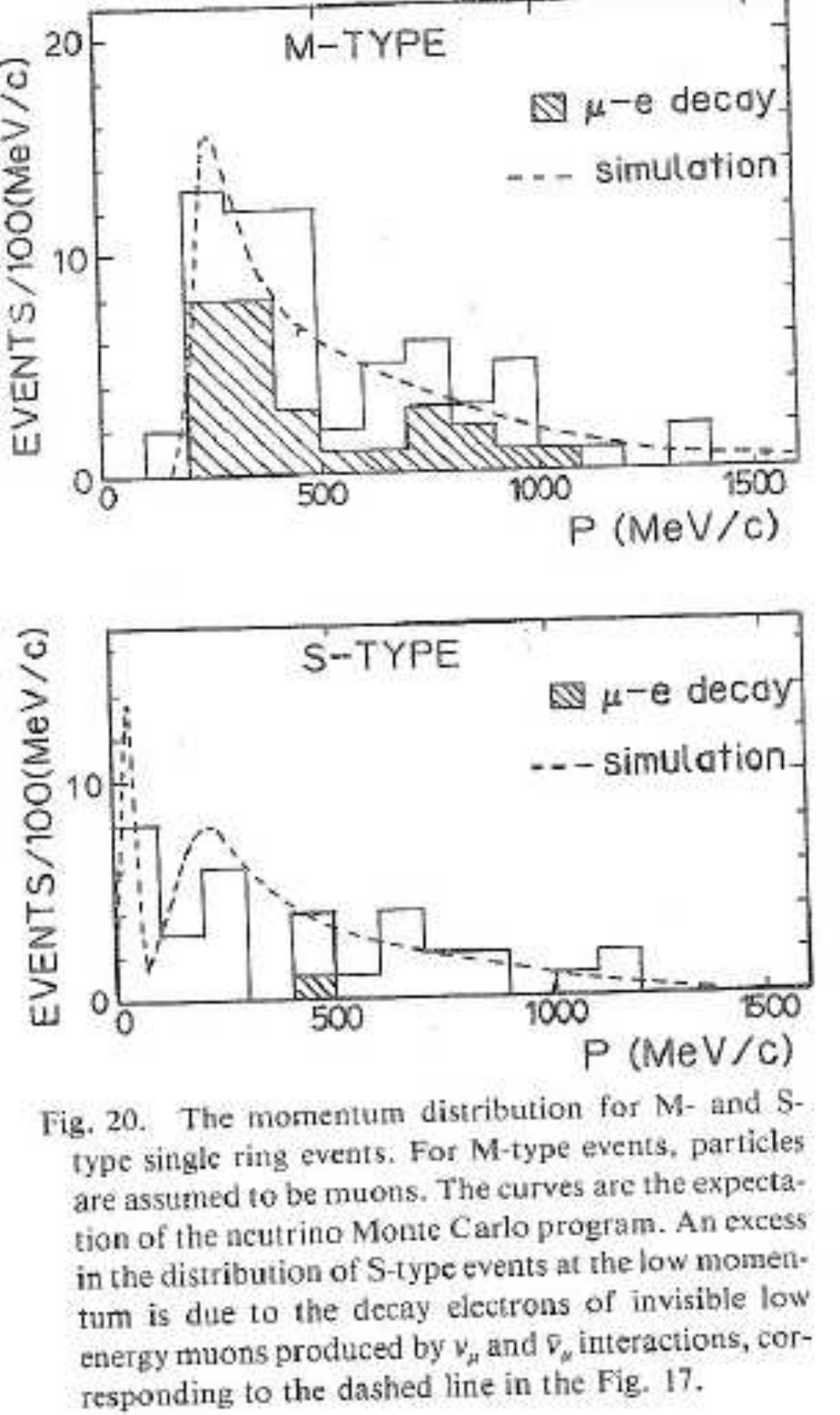}
\caption{\label{fig:NK2}The atmospheric neutrino spectrum observed with
Kamiokande I.
The upper curve illustrates the good agreement for muon type events.  The
lower figure is for electron like events.}
\end{figure}
One source of discomfort was that several other observations of atmospheric
neutrinos reported no deficit of muon events.
For example, the Kamioka equivalent\cite{Nakahata} of the IMB anomaly paper,
submitted at about the same time emphasized the good agreement of observations
with expectations (figure \ref{fig:NK1}).  ``Note the comparison is
absolute. i.e. no normalization has
been made.''  This publication summarized the Kamioka I data sample.  It
illustrated good agreement for both S, showering type events and for M, muon
type events (figure \ref{fig:NK2}).  If anything, the data presented had a
modest excess of M type
events.

The Kamioka detector had much more light collection capability
than the IMB detector.  This permitted them to utilize the shape of the
Cherenkov image to determine if the interaction had produced a muon, M type
events, or an electron, S type events.  The S stands for showering since the
electron would multiple scatter, bremsstrahlung and pair produce; a processes
known as an electromagnetic shower.  Muon induced events had a much crisper,
cleaner image.  Kamioka had used this difference in images to distinguish
electron from muon type events.

While Nakahata {\em et al.}\cite{Nakahata} had no numbers for data,
the data was the same as the Kajita PhD. thesis\cite{Kajita} from earlier in
1986.  The figures containing experimental observations are idenitical.
Kajita's PhD. thesis\cite{Kajita} reported the observations from the first
phase of Kamioka, known as Kamiokande I.  In an exposure of 1.11 kt-yr they
reported 141 contained event in 474 days of live time.  The ability to
distinguish showering from non-showering tracks permitted them to report
the event rates in the two categories.

(A kiloton-year is a measure of sensitivity to proton decay or atmospheric
neutrinos.  It means every nucleon in 1000 metric tons was observed for a
year.  One would expect twice as many interactions in two kiloton-years as in
one kiloton-year.  This could be accomplished by observing twice the mass or
by observing the same one kiloton for two years.  The neutrino flux does have
some modest time variation, due to the effect of the sun's cycle on the Earth's
magnetic field.  Also atmospheric neutrino fluxes are not uniform.  Again due
to the Earth's magnetic field there is some local variation of the flux from
place to place on Earth.  So a kiloton-year at one place will not yield the
same neutrino interactions as at another location.)

Kamiokande reported 97 single prong events (89 with energies above 100 MeV)
when they expected to observe 94 (85 with energies above 100 MeV).
So the reported event rate was as expected.  They reported 64 M type,
or muon type, events when 54 were expected.  They reported 33 S type events,
electrons (25 above 100 MeV) when 40 were expected (31 above 100 MeV).
The reason for mentioning the S type event rate above 100 MeV is that
muon decay from sub threshold muons could look like showering low energy
electron events.  Cosmic ray sub threshold muons could slip into the device
undetected and look like low energy electron neutrino interactions.

The Kamiokande I detector was capable of recording delayed muon decays
associated with an event.  29 events had muon decays when 39.3 were expected.

These numbers are all in the thesis\cite{Kajita}.  The thesis conclusions
are that the muon and electron fractions are {\em as expected}.  ``These figures
indicate that the agreement between the data and the simulation is quite
well.''\cite{Kajita}

But the numbers quoted above clearly indicate a 2.4 standard deviation
deficiency of muon decay signals and a 1.6 standard deviation excess
of M type events, when compared to expectations.  None of these significances
is calculated in the thesis.  I realized that the low muon decay rate
reported, but not noted by Kamiokande could provide confirmation of the IMB-1
3.5 standard deviation observation.

In June of 1986 I visited Tokyo following the neutrino meeting at Sendai and
met with Koshiba and Kajita.  I was well received.  Koshiba took me and a
small contingent of his group to a nearby noodle restaurant for a lesson in
the art of slurping.  The IMB anomaly paper had recently been submitted for
publication.  I discussed our observed muon decay deficiency.  I pointed out
the discrepancy between the M/S analysis and the muon decay rates in the
Kamiokande analyses.  The response was kind, blank stares.  I was assured
that the M/S analysis was correct.

Kajita's thesis\cite{Kajita} and the Nakahata paper\cite{Nakahata} were not
unique.  All Kamiokande reports stressed how closely the neutrino observations
matched expectations (table \ref{KamH}).

\section{Confirmation}
\begin{figure}
\includegraphics[height=4.5in,angle=270]{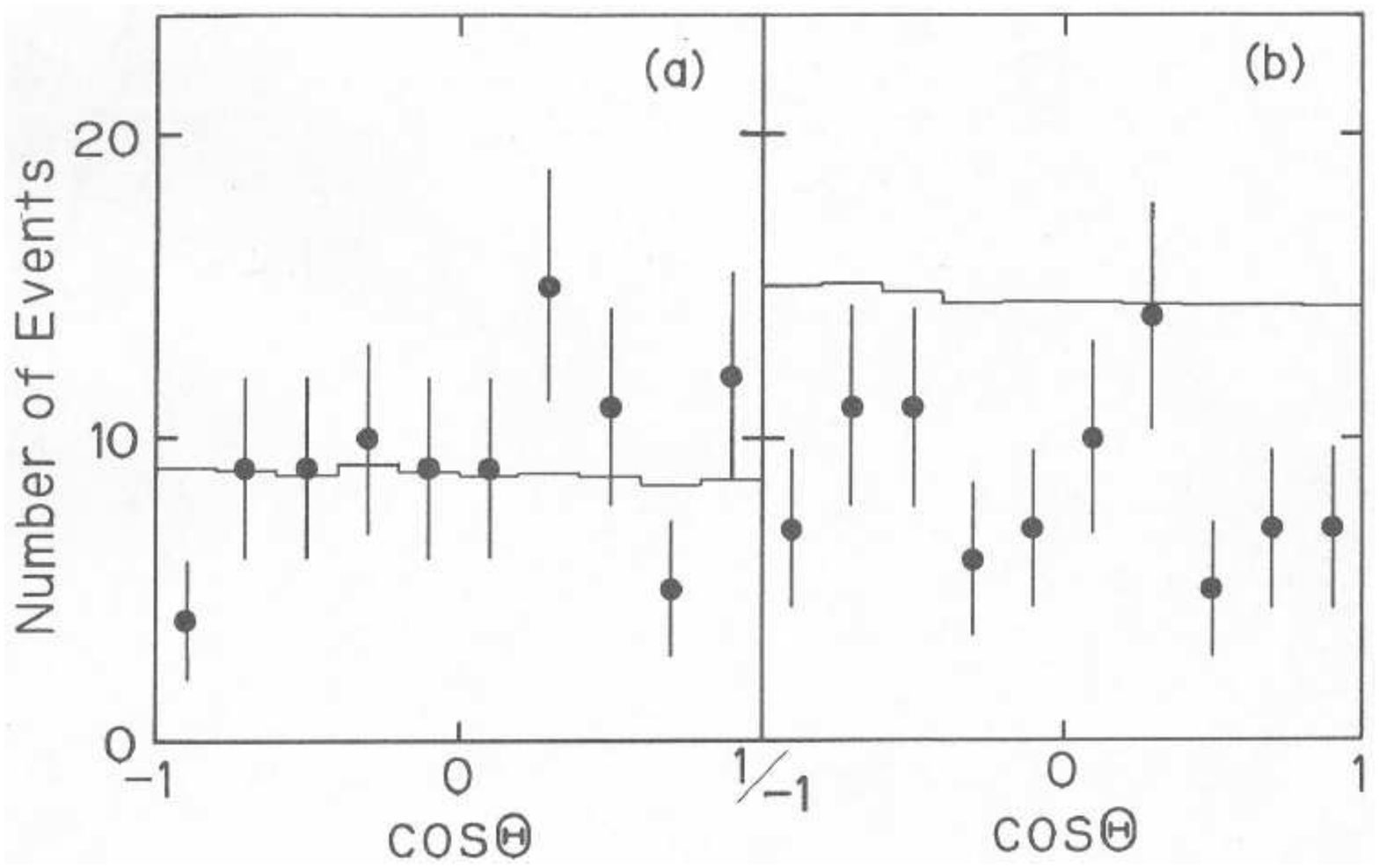}
\caption{\label{fig:Hirat}The angular distribution for electron
neutrinos (left) and muon
neutrinos (right) reported by Kamiokande in their 1988 paper confirming the
atmospheric neutrino anomaly.  Note that cosine of one is for {\em downward}
going events.  The original caption to this figure reads ``Zenith angle
distributions for; (a)electron-like events and (b) muon-like events.
$\cos(\theta)=1$ corresponds to downward going events.  The histograms show
the distributions expected from atmospheric neutrino interactions.''}
\end{figure}
In 1988 the Kamiokande experiment published a paper
(Hirata {\em et al.}\cite{Hirata}) confirming
the reported deficit of muon like neutrinos.  The paper was based on an
exposure to November 1987 with 277 contained events (265 above 100 MeV).
It had 2.87 kiloton-years of data, including the 1.11 kiloton-years from
Kamiokande I.  The M/S (muon and showering) pattern recognition classification
method {\em had been modified} to give agreement with the muon decay rate.  The
paper concluded that there appeared to be a muon deficiency.  Only 59\% of the
expected number of muon type events were observed.

The Hirata {\em et al.} paper cites and quotes the 1986 IMB muon decay
deficiency\cite{Anom}.

Interpretation of the data in this paper is still difficult since it also
shows no directional modulation (figure \ref{fig:Hirat}) or energy distortion.
A careful reading
does indicate that the event rate appeared to be lower than that reported
for the Kamiokande I data.

A brief review of the reports from Kamiokande from turn on to the 1988 paper
shows a significant change in interpretation in the 1988 paper.
\begin{table}[tb]
\begin{tabular}{rrrrrrr}
Source & Date & Exposure & Events & M type & Event Rate & Expected\\
& & kt-yrs & & Obs/MC & per kt-yr & Event Rate\\
\hline
5'th WGU\cite{5WGU} & 1984 & 0.485 & 80 & Agreed & 165 & Agreed\\
Arisaka Thesis\cite{AsakThs} & 1985 & 0.661 & 84 & 1.03 & 127 & 129\\
6'th WGU\cite{6WGU} & 1985 & 0.840 & 99 & 1.13 & 118 & 111\\
Kajita Thesis\cite{Kajita} & 1986 & 1.11 & 133 & 1.19 & 120 & 108\\
Hirata {\em et al.}\cite{Hirata} & 1988 & 2.87 & 265 & 0.59 &{\em 92} & 111\\
\end{tabular}
\caption{\label{KamH}History of Kamiokande atmospheric neutrino observations}
\end{table}
Table \ref{KamH} indicates that prior to 1988 the observed event rate was close
to expectations, if a bit high, as was the rate of muon (or M) type events.

Table \ref{KamH} also helps us resolve an issue as to the date of the
Kajita PhD. thesis\cite{Kajita}.  The report date was February 1986 as was the
cover, but the title page lists February 1985.  At the
6'th WGU in April 1985 Kamiokande reported on 840 ton-years of exposure, up to
January 23, 1985.  There could not have been enough data available from the
880 ton fiducial mass to get to 1.11 kiloton-years exposure of the thesis
anytime in February 1985.  So presumably the report date of 1986 is correct.

\FloatBarrier
\section{Epilogue}
\begin{figure}
\includegraphics[height=6in,angle=270]{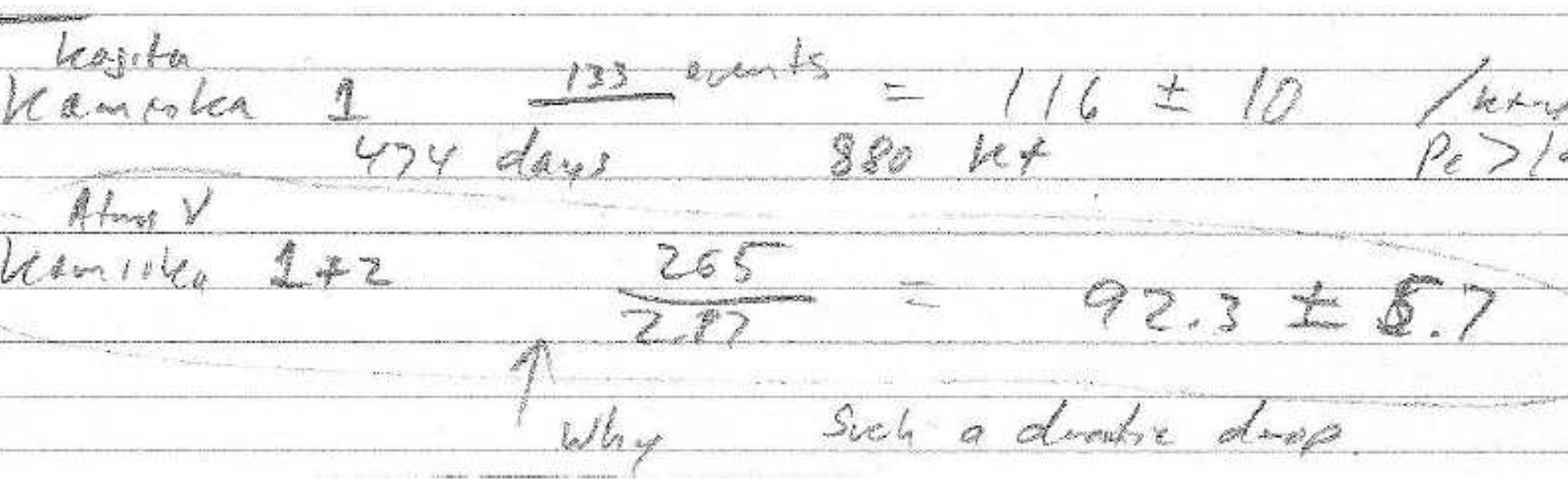}
\caption{\label{fig:Drop}Portion of handwritten notes on the event rate discrepancy between
earlier Kamiokande reports and the 1988 paper confirming the neutrino
anomaly.}
\end{figure}
The story so far has established and confirmed the observation that atmospheric
neutrinos had an apparent deficiency of muon type neutrinos.
But just as in the section {\em Scientific Context} the course of scientific
discovery, is rarely straight, there are a number of subtleties associated
with the measurements.

Personal notes written in the 1988-1989 period indicate serious concern
about the event rates reported in the Hirata {\em et al.}\cite{Hirata} paper.
Kamiokande 1 had reported an event rate of 116$\pm$10 events per kiloton-year
but the new paper based on combining this data with subsequent data
had an event rate of 92.3$\pm$5.7 per kiloton-year (figure \ref{fig:Drop}).

One can understand that the M/S numbers changed from earlier reports
because the M/S fitting method had been revised but why should the event rate
drop?  The 2.87 kiloton-year in the paper was the sum of the 1.11 kiloton-year
from Kamiokande I and 1.76 kiloton-year from Kamiokande II.  By subtraction
this means that Kamiokande II had 136 events in 1.76 kiloton-years or an
event rate of 77.3$\pm$6.8 events/ktonyr, a drop of 38\% from Kamiokande 1.

\begin{figure}
\includegraphics[height=6in,angle=270]{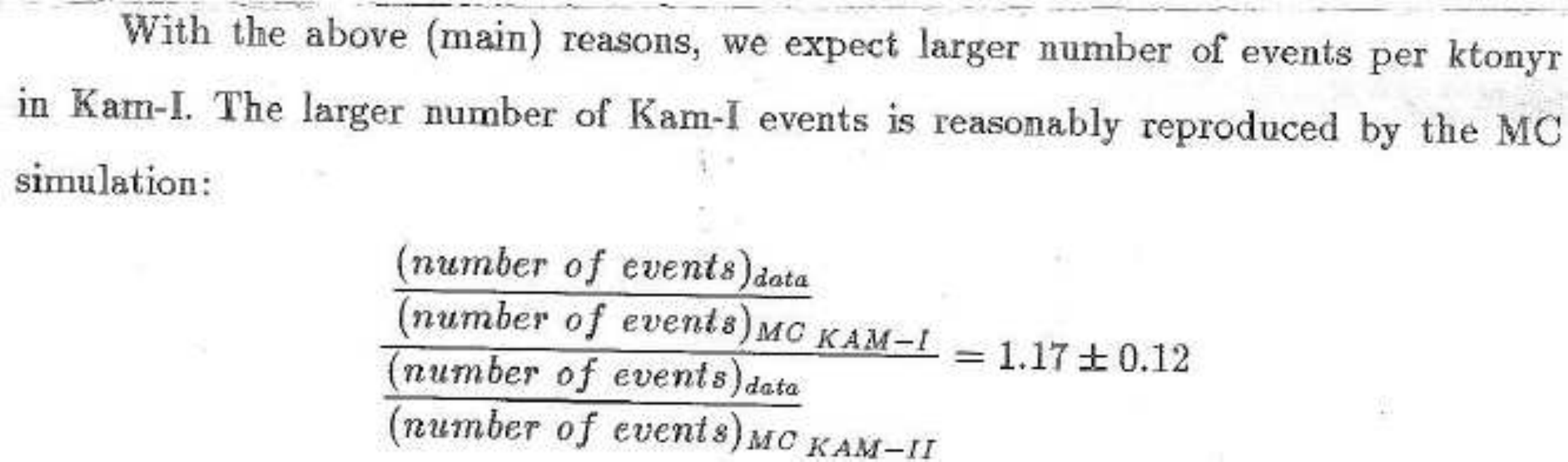}
\caption{\label{fig:KamR}Portion of the 1989 letter from the Kamiokande collaboration which
argues that the rate difference is not significant.  The 1988 paper combined
data from two detectors known as Kamiokande I and Kamiokande II.  Apparently
the event rate per kiloton-year was expected to be lower in Kamiokande II.}
\end{figure}

After discussing the rate change with Kamioka investigators at
conferences they suggested that I write to the collaboration.
An email\cite{email} was sent on April 18, 1989 and a written reply
dated August 12, 1989 came by post.
The response\cite{resp} (figure \ref{fig:KamR}) indicated that a refit of
Kamiokande I had an event
rate of 116$\pm$9.4 events/ktonyr, in good agreement with my estimates.
They also indicated that both Kamiokande I and Kamiokande II had rates
in agreement with expectations.

It would have been nice to see the muon neutrino fraction independently
for these two exposures, Kamiokande I and II.  But the data has never been
released in a format that would make that possible.


My notes indicate that a neutrino event rate check was done with the
independent IMB data sample.
IMB-1 had an event
rate of 106$\pm$5 event per kiloton-year.  In the first 1.53 kt-yr of
exposure IMB-3 had a rate of 110$\pm$10 per kiloton-year, for energies above
140 MeV.  It would appear that the neutrino flux was stable over the period
in question.

\FloatBarrier
\section{Conclusions}
The author is neither a philosopher nor an historian so readers are encouraged
to draw their own conclusions.  Scientific research is a human endeavor carried
out by people with conflicting standards.  Scientists are often expected to
draw solid conclusions from incomplete data, while maintaining an open mind.
Perhaps the best remedy for this conflict is redundancy and honest
corroboration.  In contrast to many of the issues mentioned in the introduction
the atmospheric neutrino anomaly turned out to be true and is an important
window on beyond the standard model physics.  The full story of the discovery
is more complex than can be covered in this already long article.  For example,
attempts to reconcile our measurement of a muon deficiency with results from
Nusex and Kamioka, which did not, have been left out.  Such well reasoned
attempts do not move this story forward.

In scientific research the answers are not in the back of the book and
Nature does not read Physical Review Letters.

Fred Reines, a colleague on much of the work discussed here, formulated a
poem to illustrate the challenges of observational science\cite{Reinespoem}
and is known to have recited it in the context of this research.

\begin{verbatim}
  Ode to Frustration

If at first you don't succeed
   What did you expect?
Progress would be slow indeed
   With nothing to reject.

A false step here, another there
   It means you're really trying
Besides, the struggle up the stair
   Itself is satisfying.

So labor with your charming quarks
   Though endless multiplying
And weigh each lepton, one by one,
   And look for baryons dying.

Dimuon pairs, imploding stars
   All vie for a solution
With quarks behind their prison bars
   Compounding the confusion.

Oh, Pauli, Fermi guide us
   Banish our illusions
And elevate our hunches
   To sensible conclusions.
\end{verbatim}

It took many years and various kinds of measurements to understand the
atmospheric and solar neutrino observations.  Over the course of time our
view of the significance of the results has changed.  But the story is not
over.  The neutrino sector still has some discrepant observations that
do not fit into the picture and many parameters needed to finish the picture
itself, such as the overall mass scale, are just starting to be measured.

\section{Thanks}
This work would have been impossible without the contributions of many
people.  Larry Sulak, Fred Reines, Maurice Goldhaber and Jack Van der Velde
provided the leadership for the IMB experiment, which was in many ways the
first large scale experiment in astro-particle physics.  Many creative and
hardworking people have contributed.  Most have been mentioned in the article
but should also be recognized here.  Students Bruce Cortez, Bill Foster,
Eric Shumard, Geoff Blewitt and Todd Haines all played an important role.
The next generation included Dave Casper, Steve Dye and Clark McGrew who
helped confirm the result with the IMB-3 data and some new, independent particle
classification methods.  Richard Bionta was a master at calibration and
designed and constructed the system that converted raw detector information
into useful physical observations.  Tegid W.~Jones, Danka Keilczewska and
John Learned have provided brilliant independent interpretations of the
various observations.  Many additional members of the IMB collaboration
contributed to its success.

M.~Koshiba, Y.~Totsuka and T.~Kajita have
always shown a willingness to discuss physics even if they did not always
share my point of view.

Steve Weinberg, Shelley Glashow, Howard Georgi, Abdus Salam and Jogesh Pati
provided solid theoretical motivation for the project as well as enthusiastic
support.

\section{Acknowledgments}
I would like to thank Allan Franklin for encouragement.
In particular, I benefited from many suggestions he made on an early version
of this manuscript.  The PiP editors have done a remarkable job in helping to
make the article accessible


\begin{thebibliography}{1000}
\bibitem{SouthAfr}F. Reines, M. F. Crouch, T. L. Jenkins, W. R. Kropp, H. S. Gurr, G. R. Smith J. P. F. Sellschop and B. Meyer, ``Evidence for High-Energy Cosmic-Ray Neutrino Interactions'', Phys. Rev. Lett. 15, 429-433 (1965)
\bibitem{Kolar} Krishnaswamy, M. R.; Menon, M. G. K.; Narasimham, V. S.;
Hinotani, K.; Ito, N.; Miyake, S.; Osborne, J. L.; Parsons, A. J.;
Wolfendale, A. W.``The Kolar Gold Fields Neutrino Experiment. II. Atmospheric
Muons at a Depth of 7000 hg cm-2 (Kolar)'', Proceedings of the Royal Society
of London. Series A, Mathematical and Physical Sciences, Volume 323,
Issue 1555, pp. 511-522 (1971).
\bibitem{ANeutCur}Peter Galison, ``How the first neutral-current experiments
ended'', Rev. Mod. Phys. {\bf 55}, 477509 (1983)\\
Andrew Pickering, ``Against Putting the Phenomena First: The Discovery of
the Weak Neutral Current'', Studies in the History and Philosophy of Science,
{\bf 15}, 85-117 (1984)\\
B. Aubert, A. Benvenuti, D. Cline, W. T. Ford, R. Imlay, T. Y. Ling, A. K. Mann, F. Messing, R. L. Piccioni, J. Pilcher, D. D. Reeder, C. Rubbia, R. Stefanski, and L. Sulak, `` Observation of Muonless Neutrino-Induced Inelastic Interactions'', Phys. Rev. Lett. 32, 800-803 (1974)
\bibitem{High-y}B. Aubert, A. Benvenuti, D. Cline, W. T. Ford, R. Imlay, T. Y. Ling, A. K. Mann, F. Messing, J. Pilcher, D. D. Reeder, C. Rubbia, R. Stefanski, and L. Sulak, ``Scaling-Variable Distributions in High-Energy Inelastic Neutrino Interactions'', Phys. Rev. Lett. 33, 984987 (1974)\\
A. Benvenuti, D. Cline, W. T. Ford, R. Imlay, T. Y. Ling, A. K. Mann, F. Messing, D. D. Reeder, C. Rubbia, R. Stefanski, and L. Sulak, ``Invariant-Mass Distributions from Inelastic nu and nu -bar Interactions'', Phys. Rev. Lett. 34, 597600 (1975)\\
A. Benvenuti, D. Cline, W. T. Ford, R. Imlay, T. Y. Ling,
A. K. Mann, D. D. Reeder, C. Rubbia, R. Stefanski, L. Sulak, and P. Wanderer,
``Further Data on the High-y Anomaly in Inelastic Antineutrino Scattering'',
Phys. Rev. Lett. 36, 1478-1482 (1976)
\bibitem{AnnRev86}R.~Engfer and H.K.~Walter, ``Muon Decay Experiments'',
  Ann.~Rev.~Nucl.~Part.~Sci. 1986 :327-359.
\bibitem{Lubimov}V. A. Lubimov, E. G. Novikov, V. Z. Nozik, E. F. Tretyakov
and V. S. Kosik, ``An estimate of the e neutrino mass from the Beta-spectrum
of tritium in the valine molecule'',  Phys. Lett. B94, 266 (1980).
\bibitem{Bergkvist}K.E.~Bergkvist, Nuclear Physics {\bf B39}, 839 (1972).
\bibitem{Pasierb}E. Pasierb, H. S. Gurr, J. Lathrop-, F. Reines, and H. W. Sobel, ``Detection of Weak Neutral Current Using Fission nu -bare on Deuterons'',
Phys. Rev. Lett. 43, 9699 (1979)\\
F. Reines, H. W. Sobel, and E. Pasierb, ``Evidence for Neutrino Instability'',
Phys. Rev. Lett. 45, 13071311 (1980).
\bibitem{MikSmir}S.P.~Mikheyev, A.Yu.~Smirnov, ``Resonance enhancement of
  oscillations in matter and solar neutrino spectroscopy''. Soviet Journal of
  Nuclear Physics 42 (6): 913–917, 1985
\bibitem{Bethe}H.A.~Bethe, ``Possible explanation of the solar-neutrino
  puzzle''. Physical Review Letters 56 (12): 1305–1308, 1986
\bibitem{Pontecorvo}B.~Pontecorvo, Zh.~Eksp.~Teor.~Fiz. {\bf 53}, 1717 (1967)\\
B.~Pontecorvo, Zh.~Eksp.~Teor.~Fiz. {\bf 33}, 549 (1957)\\
B.~Pontecorvo, Zh.~Eksp.~Teor.~Fiz. {\bf 34}, 247 (1958)\\
V.~Gribov and B.~Pontecorvo, Phys.~Lett. {\bf 28B} 495 (1969).
\bibitem{Perl}Martin L.~Perl, ``The Discovery of the Tau Lepton and the Changes
in Elementary-Particle Physics in Forty Years'', Physics in Perspective
{\bf 6} (2004) 401-427.
\bibitem{New-Osc}A.~K.~Mann and H.~Primakoff, ``Neutrino oscillations and the
number of neutrino types'', Phys.~Rev~{\bf D15} (1977), 655-665.
\bibitem{KobMask}M.~Kobayashi and T.~Maskawa, ``CP-Violation in the
Renormalizable Theory of Weak Interaction''
Progress of Theoretical Physics (Vol. 49 No. 2 (1973), pp.652-657)
\bibitem{Cabibbo}N.~Cabibbo, ``Unitary Symmetry and Leptonic Decays'',
Phys.~Rev.~Lett. {\bf 10} 531-533 (1963).
\bibitem{Wolf}L.~Wolfenstein, ``Neutrino oscillations in matter'',
Phys.~Rev. D 17, 2369-2374 (1978).
\bibitem{AsFr}H. David Politzer , ``Reliable Perturbative Results for Strong Interactions?'', Phys. Rev. Lett. 30, 1346-1349 (1973).\\
David J. Gross and Frank Wilczek, ''Ultraviolet Behavior of Non-Abelian Gauge Theories'', Phys. Rev. Lett. 30, 1343-1346 (1973).
\bibitem{GUTS}Howard Georgi and S. L. Glashow, ``Unity of All Elementary-Particle Forces'', Phys. Rev. Lett. 32, 438441 (1974).
\bibitem{PDK}H. Georgi, H. R. Quinn, and S. Weinberg, ``Hierarchy of Interactions in Unified Gauge Theories'', Phys. Rev. Lett. 33, 451-454 (1974).
\bibitem{Cortez}B.~Cortez, ``Birth of the Large Scale Imaging Water Cherenkov
Detector'' Sulak Festschrift, Boston University, October 22, 2005.
\bibitem{SciAm} J.M.~LoSecco, F.~Reines and D.~Sinclair, ``The Search for
Proton Decay'', Scientific American, {\bf 252} pages 54-62, June 1985.
\bibitem{E704}E.~Egelman {\em et al.}, ``A Study of the Time Evolution of a
Long-Lived $\nu_{\mu}$ Beam'', submitted to Brookhaven National Laboratory,
1977.
\bibitem{IMBprop} 
M.~Goldhaber, W.~Kropp, J.~Learned, R.~March, F.~Reines, J.~Schultz,
D.~Sinclair, H.~Sobel, L.~Sulak, J.~Vander Velde, ``A Proposal to Test for
Baryon Stability to a Lifetime of 10$^{33}$ Years'', May 1979.
\bibitem{BGC}Bruce Cortez, private communication.
\bibitem{Erice80} B.~Cortez and L.~Sulak, 
in Unification of the Fundamental Particle Interactions, Proceedings of the
Erice Workshop Europhysics Meeting, edited by S.~Ferrara, J.~Ellis and
P.~van Nieuwenhuizen March 1980, pages 661-671.
\bibitem{FWOGU-LRS} L.~Sulak, First Workshop on Grand Unification
edited by P.H.~Frampton, S.L.~Glashow and A.~Yildiz April 1980, pages 163-188.
\bibitem{Bionta1982}R.~M.~Bionta, H.~S.~Park, B.~Cortez, L.~Sulak,
``Stopping Muons in the IMB Detector'',  PDK Memo 82-6, April 1, 1982.
\bibitem{Foster1983}Bill Foster, ``The Making of the 5 Years Neutrino
Background Tapes'',  PDK Memo 83-103, July 18, 1983.
\bibitem{HarvPhd}B.~Cortez and G.W.~Foster Harvard University PhD. theses 1983.
\bibitem{Shumard}E.~Shumard University of Michigan PhD. thesis 1984.
\bibitem{Blewitt}G.~Blewitt California Institute of Technology PhD. thesis
1985.
\bibitem{LLouise}J.~LoSecco, ``Physics Results from Underground Experiments'',
New Frontiers in Particle Physics, Proceedings of the Lake Louise Winter
School edited by J.M.~Cameron, B.A.~Campbell, A.N.~Kamal and F.C.~Khanna
February 1986 pages 376-392.
\bibitem{Nusex}
G.~Battistoni {\em et al.}, ``Nucleon Decay and Atmospheric Neutrinos at the
Mont Blanc Experiment'', Proceedings of the 19'th International Cosmic
Ray Conference, La Jolla (1985) volume 8 pages 271-274.
\bibitem{KoshAspen}M.~Koshiba, Talk presented at the 1986 Aspen Winter
Conference on Particle Physics 1986.
\bibitem{Haines}T.~Haines University of California at Irvine PhD. 1986.
\bibitem{Anom}T. J. Haines, R. M. Bionta, G. Blewitt, C. B. Bratton, D. Casper, R. Claus, B. G. Cortez, S. Errede, G. W. Foster, W. Gajewski, K. S. Ganezer, M. Goldhaber, T. W. Jones, D. Kielczewska, W. R. Kropp, J. G. Learned, E. Lehmann, J. M. LoSecco, J. Matthews, H. S. Park, L. R. Price, F. Reines, J. Schultz, S. Seidel, E. Shumard, D. Sinclair, H. W. Sobel, J. L. Stone, L. Sulak, R. Svoboda, J. C. van der Velde, and C. Wuest, ``Calculation of Atmospheric Neutrino-Induced Backgrounds in a Nucleon-Decay Search'', Phys. Rev. Lett. 57, 1986-1989 (1986).
\bibitem{LOE}J. M. LoSecco, R. M. Bionta, G. Blewitt, C. B. Bratton, D. Casper,
 P. Chrysicopoulou, R. Claus, B. G. Cortez, S. Errede, G. W. Foster,
W. Gajewski, K. S. Ganezer, M. Goldhaber, T. J. Haines, T. W. Jones,
D. Kielczewska, W. R. Kropp, J. G. Learned, E. Lehmann, H. S. Park,
F. Reines, J. Schultz, S. Seidel, E. Shumard, D. Sinclair, H. W. Sobel,
J. L. Stone, L. Sulak, R. Svoboda, J. C. Vander Velde, and C. Wuest, ``A Study
 of Atmospheric Neutrinos with the IMB Detector'', Proceedings of the 19'th
International Cosmic Ray Conference, La Jolla (1985) volume 8 pages 116-119.
\bibitem{LOE3}J.~LoSecco and J.~Learned, ``Recent Results from IMB'',
Proceedings of the International School on Particles and Cosmology,
Baksan Valley, USSR, edited by V.A.~Matveev, E.N.~Alexeev, V.A.~Rubikov
and I.I~Tkachev, pages 91-108 (1991).
\bibitem{NuDK}J. M. LoSecco, R. M. Bionta, G. Blewitt, C. B. Bratton, D. Casper, R. Claus, B. Cortez, S. Errede, G. Foster, W. Gajewski, K. S. Ganezer, M. Goldhaber, T. J. Haines, T. W. Jones, D. Kielczewska, W. R. Kropp, J. G. Learned, E. Lehmann, H. S. Park, F. Reines, J. Schultz, S. Seidel, E. Shumard, D. Sinclair, H. W. Sobel, J. L. Stone, L. Sulak, R. Svoboda, J. C. Van der Velde, and C. Wuest, ``Limits on the neutrino lifetime'', Phys. Rev. D 35, 2073-2076 (1987).
\bibitem{Matter}J. M. LoSecco, ``Off-diagonal neutral currents'',
Phys. Rev. D 35, 1716-1718 (1987).
\bibitem{NuOsc}J. M. LoSecco, R. M. Bionta, G. Blewitt, C. B. Bratton, D. Casper, P. Chrysicopoulou, R. Claus, B. G. Cortez, S. Errede, G. W. Foster, W. Gajewski, K. S. Ganezer, M. Goldhaber, T. J. Haines, T. W. Jones, D. Kielczewska, W. R. Kropp, J. G. Learned, E. Lehmann, H. S. Park, F. Reines, J. Schultz, S. Seidel, E. Shumard, D. Sinclair, H. W. Sobel, J. L. Stone, L. Sulak, R. Svoboda, J. C. Vander Velde, and C. Wuest, ``Test of Neutrino Oscillations Using Atmospheric Neutrinos'', Phys. Rev. Lett. 54, 2299-2301 (1985).
\bibitem{Nakahata}M.~Nakahata {\em et al.}, ``Atmospheric neutrino background and pion nuclear effect for Kamioka nucleon decay experiment'',
 J.~Phys.~Soc.~Japan 55 (1986) 3786.
\bibitem{5WGU}M.~Koshiba, ``Results from the Kamioka Nucleon Decay Experiment''
Proceedings of the Fifth Workshop on Grand Unification, pages 13-30, edited by
K.~Kang, H.~Fried and P.~Frampton (World Scientific, 1984).
\bibitem{AsakThs}Katsushi Arisaka, ``Experimental Search for Nucleon Decay'',
PhD thesis University of Tokyo, January 1985, UTICEPP-85-01.
\bibitem{6WGU}M.~Koshiba, ``Kamioka Nucleon Decay Experiment'', Proceedings of
the Sixth Workshop on Grand Unification, pages 65-88, edited by S.~Rudaz
and T.~Walsh (World Scientific 1985).
\bibitem{Kajita}T.~Kajita PhD. thesis Tokyo University 1986.
UTICEPP-86-03 Feb. 1986.
\bibitem{Hirata}K.S. Hirata {\em et al.}, ``Experimental study of the
atmospheric neutrino flux'', Phys. Lett.  B205,(1988) 416.
\bibitem{email}J.~LoSecco Email to Totsuka April 18, 1989
\bibitem{resp}Y.~Totsuka and T.~Kajita, letter August 12, 1989.
\bibitem{Reinespoem}Fred Reines, private communication, used with permission.
\end{thebibliography}
\end{document}